\documentclass[reprint,twocolumn,superscriptaddress,amsmath,amssymb]{revtex4} 

\usepackage[dvipdfmx,hiresbb]{graphicx}
\usepackage{dcolumn}
\usepackage{bm}
\usepackage[dvipdfmx]{color}

\newtheorem{lemma}{Lemma}

\renewcommand{\Lambda}{N}
\newcommand{\eq}{Eq.}

\begin{document}

\title{
Exactly solvable model for a velocity jump\\
observed in crack propagation in viscoelastic solids
}
\author{Naoyuki~Sakumichi}
\affiliation{Soft Matter Center, Ochanomizu University, Bunkyo-ku, Tokyo 112-8610, Japan}
\author{Ko~Okumura}
\affiliation{Soft Matter Center, Ochanomizu University, Bunkyo-ku, Tokyo 112-8610, Japan}
\affiliation{Department of Physics, Ochanomizu University, Bunkyo-ku, Tokyo 112-8610, Japan}
\date{\today}

\maketitle

\noindent\textbf{
Needs to impart appropriate elasticity and high toughness to viscoelastic polymer materials are ubiquitous in industries such as concerning automobiles and medical devices. 
One of the major problems to overcome for toughening is catastrophic failure linked to a velocity jump, i.e., a sharp transition in the velocity of crack propagation occurred in a narrow range of the applied load.
However, its physical origin has remained an enigma despite previous studies over 35 years.
Here, we propose an exactly solvable model that exhibits the velocity jump incorporating linear viscoelasticity with a cutoff length for a continuum description.
With the exact solution, we elucidate the physical origin of the velocity jump: it emerges from a dynamic glass transition in the vicinity of the propagating crack tip.
We further quantify the velocity jump together with slow- and fast-velocity regimes of crack propagation, which would stimulate the development of tough polymer materials.}

\begin{figure}[b!]
 \centering
\begin{picture}(200,266)
\includegraphics[width=7cm,clip]{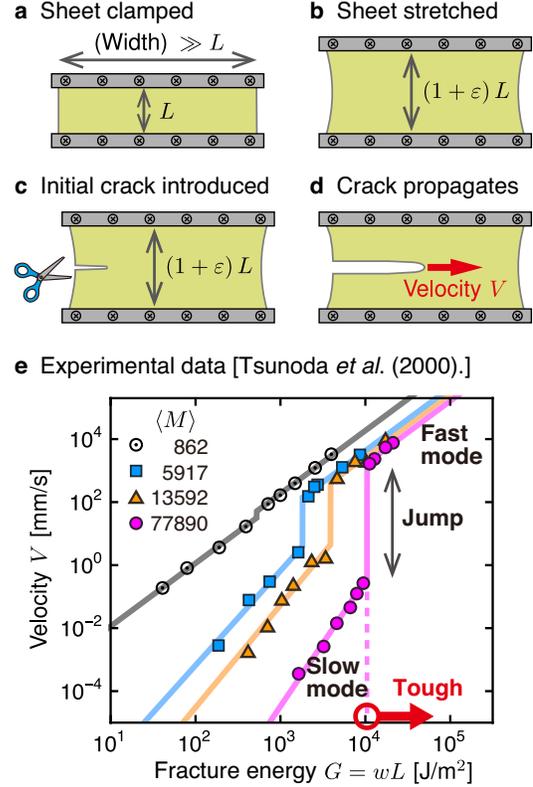}
\end{picture}
 \caption{
\textbf{
Velocity jump observed in the fixed-grip crack propagation.}
\textbf{a-d}, Schematic illustrations of the fixed-grip crack propagation investigated in the present study. 
To achieve a constant-velocity crack propagation, we perform the following four steps:
(a) we clamp the top and bottom edges of the sheet of height $L$;
(b) we stretch the sheet to a fixed strain $\varepsilon$;
(c) we introduce a small crack to initiate crack propagation; 
(d) after a short transient time, the crack propagates at a constant velocity $V$ under the fixed strain $\varepsilon$. 
In the fixed-grip crack propagation, 
the fracture energy $G$ and the strain energy release rate, 
which is expressed as $wL$ under the fixed-grip condition,
take the same value: $G=wL$.
Here, $w$ is the initially applied energy density.
\textbf{e}, Typical experimental results, $G$~vs.~$V$, obtained from the fixed-grip crack propagation by using  elastomers filled with carbon black particles (taken from Ref.~\cite{Tsunoda2000}).
With increase in fracture energy, the slow-velocity regime (a straight line on the low-velocity side) is terminated by an abrupt velocity jump, after which follows the fast-velocity regime (a straight line on the high-velocity side). 
Here, $\left<M\right>$ represent the average molar mass between nearest cross-links.
In this series of experiments, a systematic increase in the cross-link distance leads to increase in the transition energy.
Toughening is achieved by increasing the fracture energy at the transition point.
}
\end{figure}

\begin{figure*}[t]
 \centering
  \includegraphics[width=16.4cm,clip]{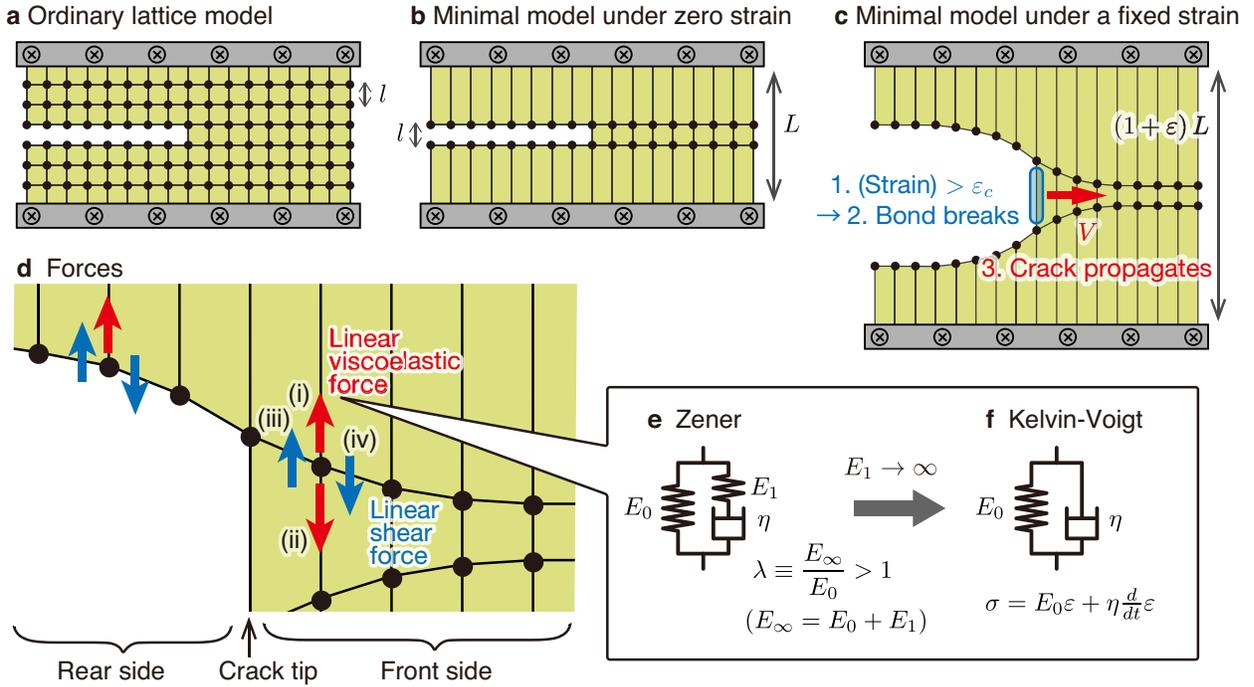}
 \caption{
\textbf{Minimal model for the fixed-grip crack propagation.}
\textbf{a}, Two-dimensional square-lattice model of a sheet with a line crack.
Here, $l$ is the lattice spacing. 
Each lattice point interacts with the nearest-neighbor points. 
We introduce a line crack by cutting bonds, i.e., 
we set the interactions of the corresponding bonds to zero.
\textbf{b}, Minimal model obtained by coarse-graining the lattice in (a).
We decimate all the lattice points except for the points on the two horizontal lines on which the two surfaces of the line crack are positioned, where $L$ is the height of the sheet under zero strain.
\textbf{c}, Mechanism of the crack propagation. When the
spring at the crack tip (encircled by a blue ellipse) is stretched to the
critical strain $\varepsilon_{c}$, the bond at the tip breaks, and, after a
certain time, the next bond at the tip is stretched to $\varepsilon_{c}$. This
cycle continues during the crack propagation. 
\textbf{d}, Forces acting on a lattice point. 
On each point in (c) located at the front side of the crack tip, four forces act: 
(i) one from the top boundary, 
(ii) one from the point below, and 
(iii, iv) the remaining two reflecting shear and acting from the left and right nearest-neighbor points. 
For each point located at the rear side, one force from the point below is missing. 
\textbf{e}, Zener element.
This element is a parallel connection of two components: 
a spring (the elastic modulus $E_{0}$) and a Maxwell element, i.e., a serial connection of another
spring (the modulus $E_{1}$) and a dashpot (the viscosity $\eta$). 
\textbf{f}, Kelvin-Voigt element obtained from a Zener element in the large $E_{1}$ limit, in which $\lambda\equiv E_{\infty}/E_{0} \to \infty$.
 }
\end{figure*}

\section*{Introduction}

Polymer-based viscoelastic materials are characterized by two elastic moduli $E_{0}$ and $E_{\infty}$ corresponding to (soft) rubbery and (hard) glassy states, respectively
\cite{DoiEdwards1988, ferry1980viscoelastic}. 
From this standard picture, one can understand generic features of the dependence of fracture energy on the velocity of crack propagation \cite{Gent1996Langmuir,Kashima2014}: 
the fracture energy $G$ 
(twice the energy required to create a crack surface of unit area \cite{Anderson})
starts from a static value $G_{0}$ and increases with the velocity $V$ to the
value $\lambda G_{0}$ with the ratio $\lambda \equiv E_{\infty}/E_{0}$ ($\simeq10^{2}$--$10^{3}$)
\cite{PGG1996softAdhesive,SoftInterface,saulnier2004adhesion}. 
This is because strong dissipation occurs at places far from the crack tip, 
whereas $G_{0}$ is well described by the cutting energy of chemical bonds and an effective cross-link distance \cite{LakeThomas1967}.

To further investigate dynamic properties of $G$ as a function of $V$, 
crack propagation experiment performed under a fixed-grip (or pure-shear) condition possesses significant advantages. 
We illustrate this experiment in Fig.~1a-d: 
a long sheet of height $L$ is subject to a fixed strain $\varepsilon$ before and after the initiation of crack propagation,
unlike other experiments based on peeling, tearing, cyclic loads, etc. \cite{lake2003fracture,Thomas2002RubberChemTech}. 
Advantages of the fixed-grip experiment are also stressed in Ref.~\cite{Langer1989PRA}, and here we emphasize the following two points. 
(i) A steady-state crack propagation is realized with no work done by the external force, 
which leads to the equality $G=wL$ with the initially applied elastic energy density 
 $w\equiv\int_{0}^{\varepsilon}\sigma(\epsilon)d\epsilon$, where $\sigma$ is the stress \cite{lake2003fracture,RivlinThomas1953,Morishita2016PRE,Morishita2017Polymer}.
(ii) The experiment shown in Figure~1e \cite{Tsunoda2000} and many other experiments \cite{Thomas1981RubberFrac,Morishita2016PRE,Morishita2017Polymer} indicate that the $G$-$V$ plots exhibit an intriguing structure for elastomers: 
the velocity $V$ jumps at a critical value $G=G_{c}$, causing a transition from the slow-velocity ($V\lesssim1$ mm/s) to fast-velocity ($V\gtrsim10^{3}$ mm/s) regime.
This $G$-$V$ structure reveals that toughness is achieved by increasing the critical value $G_{c}$ because such an increase reduces the risk of a velocity jump, which can trigger catastrophic failure. 

Theoretical understanding of the velocity jump has been very limited, although it is important for toughening polymer materials. 
Previous theories based on linear fracture mechanics \cite{Anderson} and linear viscoelasticity \cite{ferry1980viscoelastic} are unable to reproduce the velocity jump \cite{GreenwoodJohnsonRate,Langer1989PRA,Persson2005PRE}.
Although there is a theory that reproduces the jump \cite{PerssonPRL2005}, the theory predicts an extremely high-temperature region near the crack tip whereas only a slight temperature-increase was experimentally observed \cite{GaliettiHotCrack2013}.

In this article, we propose a minimal model that exhibits the velocity jump observed in the fixed-grip crack propagation, incorporating linear viscoelasticity with using the two elastic moduli $E_{0}$ and $E_{\infty}$.
This is performed with a spirit similar to the ones with which one of the authors constructed simple
and useful models for biological composites
\cite{OkumuraPGG2001,AoyanagiOkumura2010,OkumuraMRS2015}.
From the proposed model, we obtain successfully an exact analytical relation between the initially applied energy density $w$ and the crack propagation velocity $V$.  
As a result, we find simple expressions characterizing the transition point.
These expressions provide guiding principles to reduce the risk of the jump, which can trigger catastrophic failure.
Furthermore, we elucidate the physical mechanism that leads to the jump, indicating a direct link to dynamic emergence of a glassy state at the crack tip, 
and our results imply that the jump could be universally observed in a broad class of viscoelastic materials in addition to elastomers. 

\section*{Results}

\subsection{Minimal model that exhibits the velocity jump}

To construct the minimal model, we start from the two-dimensional square-lattice model (Fig.~2a), often used to simulate the structure and dynamics of sheet materials, with the lattice spacing $l$ and the sheet height $L$ under zero strain. 
Then, we derive a simplified model illustrated in Fig.~2b by decimating most of the lattice points.
As shown in Fig. 2c, the survivors (lattice points) represent the minimum number of variables essential to describe crack propagation.
To realize a crack propagation in the $x$-direction (i.e., horizontal direction), we assume each bond to break if the local strain at the crack tip is larger than the critical strain $\varepsilon_{c}$. 
For simplicity, we assume that the sheet is symmetric about the $x$-axis, and thus we consider only the lattice points on the upper side.

We explain the forces acting on each remaining lattice point on the upper side illustrated in Fig.~2d.
We assume that Poisson's ratio is zero. 
Thus, the forces always orient towards the $y$-direction (i.e., vertical direction) and
each point can move only in the $y$-direction (see Supplementary Section~I for details).
Let $u_{i}$ be the $y$-coordinate of the $i$-th point. 
The equation of motion of lattice points in the $y$-direction is given by
\begin{equation}
m\frac{\partial^2}{\partial t^2} u_{i}
= K(u_{i}-u_{i+1}+u_{i}-u_{i-1})+F_i,
\label{eq:EoM}
\end{equation}
where $K(u_{i}-u_{i+1}+u_{i}-u_{i-1})$ represents linear-elastic shear force acting from the left and right nearest-neighbor points,
and $F_i$ represents viscoelastic tensile force acting from the top boundary and the point below.
The tensile force $F_i$ is described by a Zener element in Fig.~2e characterized by two elastic moduli ($E_{0}$ and $E_{\infty}$) and viscous dissipation ($\eta$), as in de Gennes' trumpet model \cite{PGG1996softAdhesive,SoftInterface,saulnier2004adhesion}.
As illustrated in Fig.~2d, $F_i$ takes two different forms, depending on whether the $i$-th lattice point is located on the rear (i.e., left) or front (i.e., right) side of the crack tip 
because one of the four forces is missing on the rear side.
We relegate the explicit form of $F_i$ to Supplementary Section~III to avoid complication.
Instead, we give the explicit form of $F_i$ in the limit, $E_{\infty} \rightarrow\infty$, in which a Zener element reduces to a Kelvin-Voigt element (Fig.~2f).
In this limit, the tensile forces on the rear and front sides take the following form:
\begin{equation}
F_i= 
\begin{cases}
\alpha\left[c-l\left(E_{0}u_{i}+\eta\frac{\partial}{\partial t}u_{i} \right)\right] & \textrm{rear side} \\
\alpha\left[c-L\left(E_{0}u_{i}+\eta\frac{\partial}{\partial t}u_{i} \right)\right] & \textrm{front side},
\end{cases}
\label{eq:Voigt}
\end{equation}
where $\alpha$ and $c$ are constants.

\begin{figure}[t!]
 \centering
\includegraphics[width=7.2cm,clip]{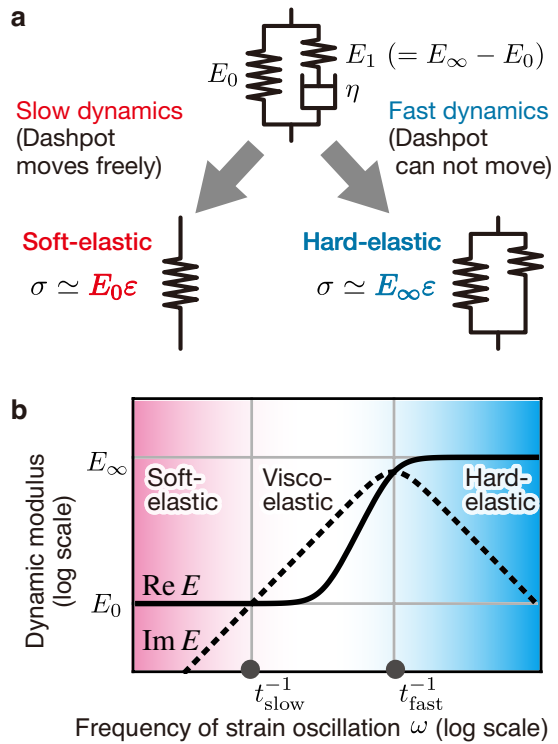}
 \caption{
\textbf{Three types of dynamic responses of a Zener element: soft-elastic, viscoelastic, and hard-elastic responses.}
\textbf{a}, 
Dynamical response of a Zener element to adequately slow or fast stretch (see, text).
\textbf{b}, 
Dynamic modulus $E(\omega)=\mathrm{Re}E(\omega)+i\,\mathrm{Im}E(\omega)$ as a function of an angular frequency of strain oscillation $\omega$ in a Zener element.
Here, $\mathrm{Re}E(\omega)$ and $\mathrm{Im}E(\omega)$ are the storage and loss moduli, respectively.
We plot $E(\omega)=E_{0}(1+i\omega t_{\mathrm{slow}})/(1+i\omega t_{\mathrm{fast}})$ obtained from the stress-strain relation in \eq~(\ref{eq:stress-strain}),
with $t_{\mathrm{fast}}\simeq\eta/E_{\infty}$ and 
$t_{\mathrm{slow}}\simeq\eta/E_{0}$. 
In the (rubbery) soft-elastic and (glassy) hard-elastic regimes, 
the dynamics are elastic and characterized by $E_{0}$ and $E_{\infty}$, respectively, 
whereas in the viscoelastic regime the dynamics is governed by $\eta$, $E_{0}$, and $E_{\infty}$.
}
\end{figure}

\begin{figure*}[t!]
 \centering
\includegraphics[keepaspectratio, scale=1.0]{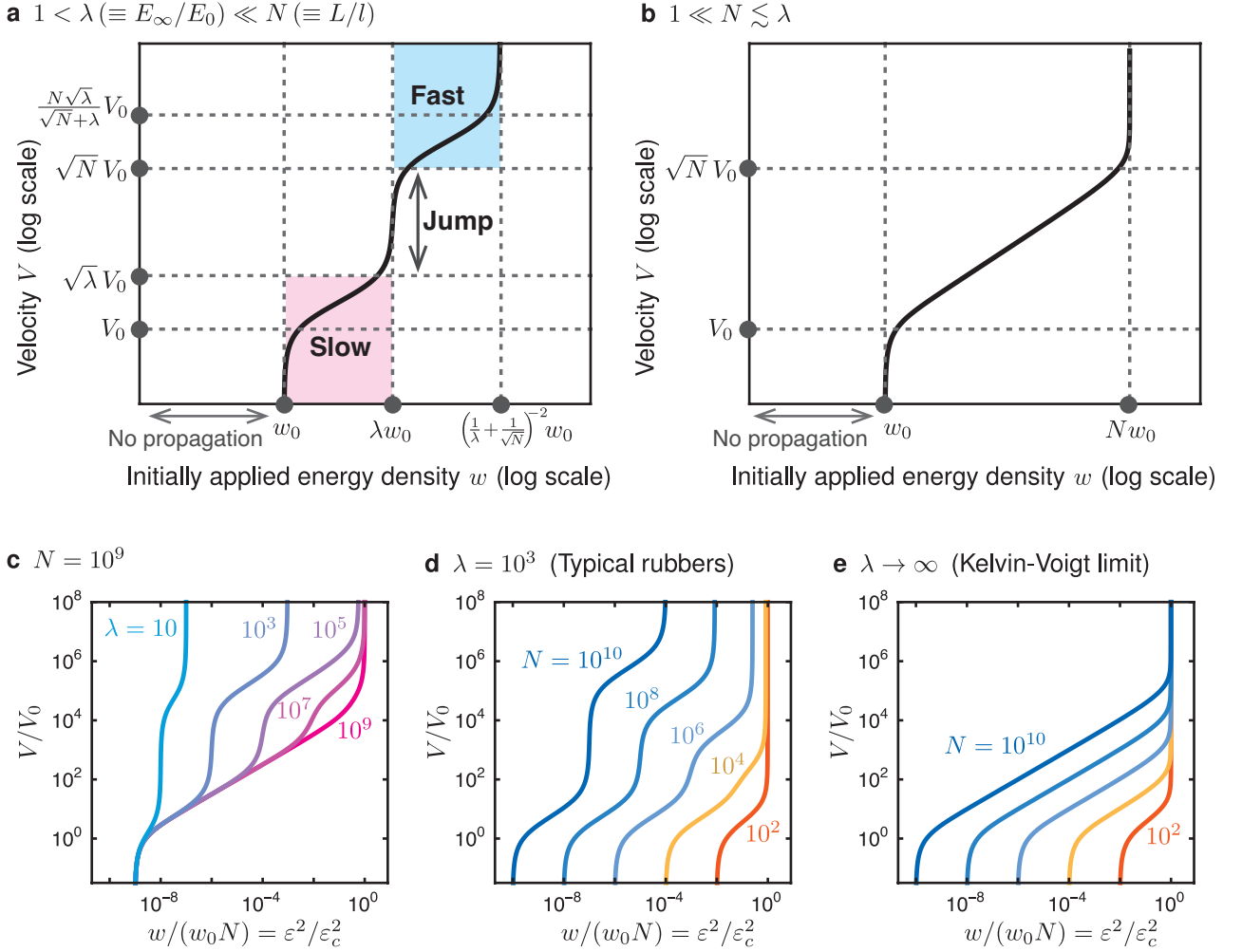}
 \caption{
\textbf{
Reproduced velocity jump and simple characterization of the $w$-$V$ curve.}
\textbf{a-b}, Two representative plots of the crack propagation velocity $V$
as a function of the initially applied energy density $w$.
The cases (a) with and (b) without velocity jump are obtained for 
$1<\lambda\ll\Lambda$ and
$1\ll\Lambda\lesssim\lambda$, respectively.
(These plots are obtained for (a) $\lambda=10^3$, $\Lambda=10^9$ and (b) $\lambda\to\infty$, $\Lambda =10^9$.)
Four characteristic velocity-scales and three energy-scales are
indicated in (a), which are important for toughening.
\textbf{c-e}, $V/V_{0}$ vs. $w/(w_{0}\Lambda)=\varepsilon^{2}/\varepsilon_{c}^{2}$,
obtained from \eq~(1) on a log-log scale. 
The normalization factors for velocity and energy are $V_{0}\simeq lE_{0}/\eta$ and $w_{0}\equiv
E_{0}l\varepsilon_{c}^{2}/(2L)$, respectively. 
The cases with velocity jump are demonstrated for various $\lambda$ with a fixed $\Lambda$ in (c)
and for various $\Lambda$ with a fixed $\lambda$ in (d).
The Kelvin-Voigt limit, $\lambda\rightarrow\infty$, is shown for various $\Lambda$ in (e) as an example of
the case without velocity jump.
}
\end{figure*}

A Zener element \cite{Tschoegl,MainardiSpada2011,Chyasnavichyus2014} is one of the simplest models to represent a typical viscoelastic behavior around a glass transition for polymer materials.
As illustrated in Fig.~3a, when stretched with an adequately slow speed, a Zener element exhibits a (rubbery) soft-elastic behavior, because the dashpot moves freely without any friction: the elastic modulus is small and approximately given by $E_{0}$.
On the other hand, when stretched with an adequately fast speed, a Zener element exhibits a (glassy) hard-elastic behavior, because the dashpot does not have enough time to move: 
the elastic modulus is large and approximately given by $E_{\infty}$.
For a conventional elastomer, $\lambda\equiv E_{\infty}/E_{0} \simeq 10^3$.
The relation between stress ($\sigma$) and strain ($\mathcal{E}$) of Zener element is given by
\begin{equation}
\left(1+t_{\mathrm{fast}}\frac{d}{dt}\right)\sigma (t)
=\left(1+t_{\mathrm{slow}}\frac{d}{dt}\right)
E_{0}\mathcal{E}(t),
\label{eq:stress-strain}
\end{equation}
with $t_{\mathrm{fast}}\equiv\eta/E_{1}\simeq\eta/E_{\infty}$ and 
$t_{\mathrm{slow}}\equiv\eta/E_{0}+\eta/E_{1}\simeq\eta/E_{0}$. 
As shown in Fig.~3b, \eq~(\ref{eq:stress-strain}) gives a dynamic modulus (i.e., the ratio of stress to strain under oscillatory conditions), mimicking a typical viscoelastic behavior around a glass transition for polymer materials.

\subsection{Exact analytical relation between $w$ and $V$}

The minimal model allows us to derive an exact analytical relationship between $w=G/L$ and $V$, 
in a continuum limit in the $x$-direction, in which we replace $u_{i}-u_{i+1}+u_{i}-u_{i-1}$ with $l^{2}\partial^{2}u(x,t)/\partial x^{2}$ in \eq~(\ref{eq:EoM}).
For simplicity, we further take the overdamped (i.e., inertialess) limit, i.e., we neglect the inertial term $m\partial^{2}u/\partial t^{2}$.
The latter limit is valid if the crack propagation velocity under question is much smaller than the shear wave velocity $l\sqrt{K/m}$.
Under the two limits, we rewrite \eq~(\ref{eq:EoM}) as
\begin{equation}
0=l^{2}K\frac{\partial^{2}}{\partial x^{2}}u(x,t) + F(x,t). 
\label{eq:EoM2}
\end{equation}
Here, the form of $F(x,t)$ changes depending on whether the position $x$ is located on the rear or front side of the crack tip as implied above, and \eq~(\ref{eq:EoM2}) satisfies appropriate boundary conditions at $x=\pm\infty$ and matching conditions at the crack tip.

We now explain the main result: an exact analytical relation between $w$ and $V$ (see Methods for the derivation).
Since the present model is initially (i.e., before the crack propagates) at rest with a fixed $\varepsilon$ without shear, 
it behaves as a linear elastic material goverfned by 
$\sigma=E_{0}\varepsilon$ and the initially applied energy density is given by $w=E_{0}\varepsilon^{2}/2$.
Let $\Lambda \equiv L/l$ be the dimensionless parameter of the length scale in the $y$-direction.
For $\varepsilon \leq \varepsilon_c /\sqrt{\Lambda}$ the crack does not propagate ($V=0$)
and for $\varepsilon \geq \varepsilon_c \lambda/(\sqrt{\Lambda}+\lambda -1)$
any constant-velocity solutions do not exist.
(When $\varepsilon \to \varepsilon_c \lambda/(\sqrt{\Lambda}+\lambda -1)$, the velocity $v$ diverges to infinity, which is an artifact resulting from the overdamped limit.)
The crack propagates with a constant velocity only in the range
$\varepsilon_c /\sqrt{\Lambda} < \varepsilon <  \varepsilon_c \lambda/(\sqrt{\Lambda}+\lambda -1)$,
or equivalently, in the range 
$w_0<w<w_\infty$.
Here, $w_0\left(\equiv E_{0}\varepsilon_{c}^{2}/(2\Lambda)\right)$ and $w_\infty$ are the minimum and maximum values of $w$ for the propagation with a constant velocity, respectively.
In this range, the relation between $w$ and $V$ is given by
\begin{equation}
\frac{w}{w_{0}}=\Lambda\left[\frac{\frac{1}{\lambda-1}\left(\frac{\Lambda}{\xi_{1}\xi_{\Lambda}}+\lambda\right)  \frac{V}{V_{0}}+\xi_{1}+\xi_{\Lambda}}{\frac{1}{\lambda-1}\left(  \Lambda+\frac{\Lambda}{\xi_{1}\xi_{\Lambda}}+\lambda-1\right)  \frac{V}{V_{0}}+\Lambda\xi_{1}+\xi_{\Lambda}}\right]^{2},
\label{eq:Result-Zener}
\end{equation}
with a reference velocity
$V_{0}\equiv\frac{l}{\eta}\sqrt{\frac{1}{2}\left(1-\frac{1}{N}\right)E_{0}\mu}$,
where $\mu$ is an effective shear modulus.
We note that $V_{0}$ scales as $l/t_{0}$ with the (largest) relaxation time $t_{0}\equiv\eta/E_{0}$, in practical cases with $l\ll L$, in which $\mu$ scales as $E_{0}$.
In \eq~(\ref{eq:Result-Zener}), the dimensionless length scale $\xi_{\Lambda}$ is the positive real solution of the following cubic equation for $\xi$:
\begin{equation}
\xi^{3}+\frac{\lambda V}{(\lambda-1)V_{0}}\xi^{2}-\Lambda\xi-\frac{\Lambda V}{(\lambda-1)V_{0}} =0,
\label{eq:CharacteristicEq}
\end{equation}
which has a unique positive real solution as guaranteed by Lemma~1 in Supplementary Section~III-B.
The explicit form of $\xi_{\Lambda}$ is given by Cardano's formula for the solution of a cubic equation \cite{Math}.
We obtain $\xi_1$ by substituting $\Lambda =1$ to $\xi_{\Lambda}$.

As illustrated in Fig.~4a, \eq~(\ref{eq:Result-Zener}) guarantees the existence of the velocity jump for $\lambda\ll \Lambda \equiv L/l$.
The existence condition $\lambda\ll \Lambda $ is derived in Supplementary Section~IV-B (see, Theorem~3)
and is well satisfied in conventional elastomers for regular specimen sizes
($\lambda\simeq10^{3}$, $L\simeq10$~cm, and $l\simeq10$~nm).
Since a Zener element generally represents a typical viscoelastic behavior around a glass transition,
the present model is relevant to a broad class of materials beyond elastomers:
the velocity jump is expected to be a universal phenomenon in polymer materials such as gels and resins.
Note that the present model does not reproduce the velocity jump 
for $\Lambda\lesssim\lambda$ (including the Kelvin-Voigt limit, $\lambda\to\infty$)
as illustrated in Fig.~4b.
Figure~4c-e demonstrates how \eq~(\ref{eq:Result-Zener}) depends on $\lambda$ and $\Lambda$.

\begin{figure}[b!]
 \centering
\includegraphics[width=8.45cm,clip]{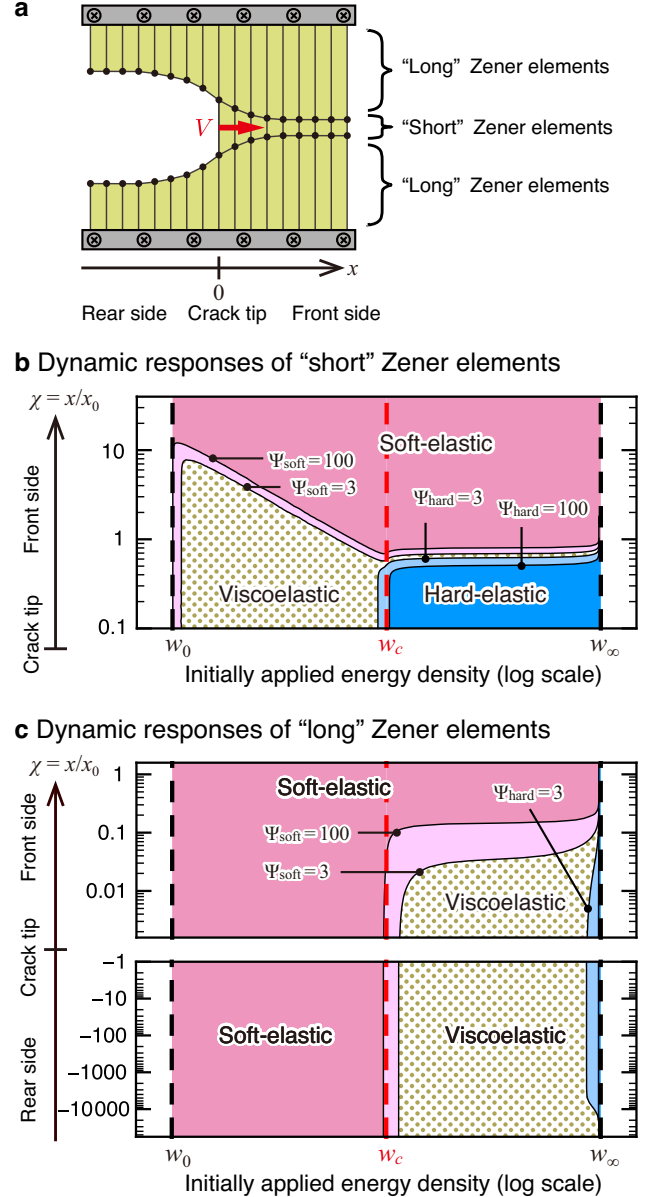}
 \caption{
\textbf{Dynamic responses of Zener elements.}
\textbf{a}, ``Short'' and ``long'' Zener elements of natural length $l$ and $(L-l)/2$, respectively.
Here, we set the $x$-coordinate origin at the crack tip.
\textbf{b-c}, Representative behavior of dynamic responses of Zener elements for conventional elastomers ($\lambda =10^3$ and $\Lambda =10^9$).
We show relaxation responses of Zener elements during a constant-velocity crack propagation, in contrast to vibration responses in Fig.~3.
The four curves are contours for $\Psi_\mathrm{soft}=100, 3$ and $\Psi_\mathrm{hard}=3, 100$ as a function of the initially applied energy density $w$ and the distance from the crack tip $\chi\equiv x/x_{0}$.
The conditions $\Psi_\mathrm{soft} \gg 1$ and $\Psi_\mathrm{hard} \gg 1$ correspond to soft- and hard-elastic regimes, respectively (see the text for details).
Explicit forms of $\Psi_\mathrm{soft}$, $\Psi_\mathrm{hard}$, and $x_{0}$ are given in Methods.
Red dashed lines correspond to the velocity jump ($w_c$) and black dashed lines correspond to the minimum ($w_0$) and maximum ($w_\infty$) values of $w$ for with constant-velocity propagation.
}
\end{figure}

\begin{figure*}[t!]
 \centering
\includegraphics[width=17.2cm,clip]{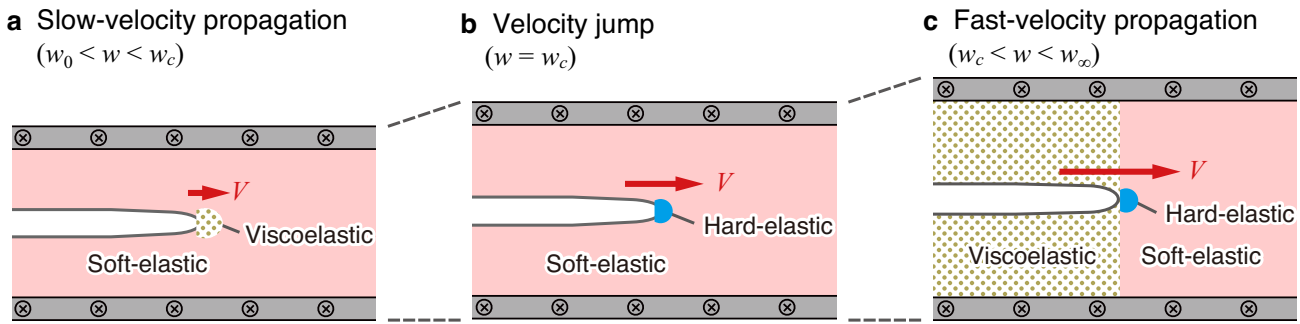}
 \caption{
\textbf{Physical pictures on the crack propagation revealed by the present exact solution.}
We draw the three illustrations based on Fig.~5.
\textbf{a}, The slow-velocity propagation is characterized by viscous dissipation in the vicinity of the crack tip. 
\textbf{b}, The velocity jump induced by emergence of a hard-elastic regime (as a result of a dynamic glass transition) in the close vicinity of the crack tip.
\textbf{c}, The fast-velocity propagation is characterized by viscous dissipation on the rear side (with the hard-elastic regime in the close vicinity of the crack tip).
Note that away from the crack tip viscous dissipation in the viscoelastic regime decays with the distance from the tip (see the text for details).
}
\end{figure*}

\subsection{Guiding principles to develop tough polymer materials}

The exact relation in \eq~(\ref{eq:Result-Zener}) leads to simple expressions for the four points characterizing the $w$-$V$ curve given in Fig.~4a, 
such as $(w_{0},V_{0})$ and $(\lambda w_{0},\sqrt{\lambda}V_{0})$.
In particular, the point $(\lambda w_{0},\sqrt{\lambda}V_{0})$ shows that the velocity jump occurs at $w=w_{c}$, where
\begin{equation}
w_{c}\equiv\lambda w_{0}=\frac{lE_{\infty}\varepsilon_{c}^{2}}{2L}.
\label{eq:TransitionEnergy}
\end{equation}
The transition energy density $w_c$ given in \eq~(\ref{eq:TransitionEnergy}) is consistent with empirical knowledge in polymer science. 
For instance, Fig.~1e experimentally shows that the transition energy $G_{c}=w_{c}L$ increases as the cross-link distance (i.e., the parameter $l$) increases \cite{Tsunoda2000}. 
This feature is consistent with \eq~(\ref{eq:TransitionEnergy})
because $E_{\infty}$ and $\varepsilon_{c}$ are approximately constant even for different $\left<M\right>$ in Fig.~1e
(see, e.g., Ref.~\cite{Morishita2016PRE}).

Equation~(\ref{eq:TransitionEnergy}) gives the following guiding principles to develop tough polymer materials
(i.e., to reduce the risk of a velocity jump, which can trigger catastrophic failure):
the transition energy density $w_c$ is enhanced with increase in
(i) the modulus $E_{\infty}$ of the glassy state and/or (ii) the lattice spacing $l$.
Here, we can regard $l$ as a characteristic length scale below which the continuum description is no longer valid: 
$l$ is the largest length scale among scales such as the cross-link distance, the size of filler particles, the filler-particle distance, and the length scale of possible inhomogeneous structures in the sample.
Equation~(\ref{eq:TransitionEnergy}) indicates that we can keep the appropriate principal elasticity $E_{0}$ to develop tough polymer materials in principle, which is a practical advantage.

\subsection{Physical origin of the velocity jump}

To elucidate the physical origin of the velocity jump, we focus on a crossover among the three types of dynamic responses of Zener elements, corresponding to soft-elastic, viscoelastic, and hard-elastic regimes (Fig.~3b), depending on the time scale of the propagation dynamics.
Since we are interested in a crack propagation closely related to relaxation responses (rather than oscillatory responses in Fig.~3b) of Zener elements, we introduce the two parameters
\begin{equation*}
\Psi_{\mathrm{soft}}\equiv \frac{\mathcal{E}}{t_{\mathrm{slow}}\frac{d}{dt}\mathcal{E}}
\quad\mathrm{and}\quad
\Psi_{\mathrm{hard}}\equiv \frac{t_{\mathrm{fast}}\frac{d}{dt}\sigma}{\sigma},
\end{equation*}
to characterize the dynamic responses behind \eq~(\ref{eq:stress-strain}):
(i) when $\Psi_{\mathrm{soft}} \gg 1$ (and $\Psi_{\mathrm{hard}} \ll 1$), 
\eq~(\ref{eq:stress-strain}) reduces to $\sigma =E_{0}\mathcal{E}$, which corresponds to the soft-elastic regime;
(ii) when $\Psi_{\mathrm{hard}} \gg 1$ (and $\Psi_{\mathrm{soft}} \ll 1$), 
\eq~(\ref{eq:stress-strain}) reduces to $\sigma = E_{\mathrm{\infty}}\mathcal{E}$ (with omission of an integral constant), which corresponds to the hard-elastic regime;
(iii) when $\Psi_{\mathrm{soft}} \lesssim 1$ and $\Psi_{\mathrm{hard}} \lesssim 1$, viscous dissipation terms in \eq~(\ref{eq:stress-strain}) play a role in the dynamics, which corresponds to the viscoelastic regime.

By using the parameters $\Psi_{\mathrm{soft}}$ and $\Psi_{\mathrm{hard}}$, 
we show in Fig.~5 dynamic responses of the ``short'' and ``long'' Zener elements (see, Fig.~5a) in the present model.
To clarify physical pictures for the slow-velocity ($w_{0}<w<w_{c}$) and fast-velocity ($w_{c}<w<w_{\infty}$) crack propagations and the velocity jump ($w=w_{c}$), we should pay attention to the moving Zener elements near the crack tip.
In other words, the Zener elements far from the crack tip are almost in equilibrium and do not affect crack-propagation dynamics.
For example, soft-elastic regimes in Fig.~5b-c are almost in equilibrium and play a minor role for crack propagations.
Thus, we now focus on the viscoelastic and hard-elastic regimes in Fig.~5b-c.
Figure 5b shows that the ``short'' Zener element in the vicinity of the crack tip is viscoelastic in the slow-velocity propagation ($w_{0}<w<w_{c}$) but is hard-elastic in the fast-velocity propagation ($w_{c}<w<w_{\infty}$), with an abrupt change at the velocity jump ($w=w_{c}$).
Figure 5c shows that the ``long'' Zener elements near the crack tip are soft-elastic and viscoelastic in slow- and fast-velocity propagations, respectively, with an abrupt change at $w=w_{c}$. 
Note that the viscoelastic regime far from the crack tip on the rear side ($-\chi \gg 1$) in Fig.~5c is almost in equilibrium and accompanied by exponentially-small viscous dissipation.
In fact, the stress ($\sigma$), strain ($\mathcal{E}$), and their time derivatives (given by \eq~(\ref{eq:solutions-long-rear}) in Method) decay with the same exponential factor as the distance from the crack tip is increased, whereas $\Psi_\mathrm{soft}$ and $\Psi_\mathrm{hard}$, by definition, take finite values even at far distances.

From the above observations, we can draw physical pictures for the slow- and fast-velocity crack propagations and the velocity jump as illustrated in Fig.~6:
(i) the slow-velocity and fast-velocity crack propagations are characterized by viscous dissipation in the vicinity of the crack tip (Fig.~5b) and on the rear side (Fig.~5c), respectively, as illustrated in Fig.~6a and c;
(ii) The velocity jump starts due to the emergence of a hard-elastic regime near ahead of the crack tip (Fig.~5b) and ends due to the emergence of a viscoelastic regime on the rear side (Fig.~5c), as illustrated in Fig.~6b. 
Since the appearance of a hard-elastic regime is a sign of the dynamic glass transition, 
we can interpret the onset of the velocity jump at $w=w_{c}$ (Fig.~6b) as the dynamic glass transition at the crack tip.
Note that the glass transition occurs practically only in the close vicinity of the crack tip because the transition requires a strong stretch and such a stretch can occur only for short elements. This fact implies that a glass transition is easy to occur in crack propagation, and thus, we expect that even materials such as gels, in which glass transitions are difficult to occur, could exhibit a velocity jump.

\section*{Discussion}

In summary, we have proposed a minimal model that exhibits the velocity jump in viscoelastic solids for which an exact analytical solution is available.
The exact relation given in \eq~(\ref{eq:Result-Zener}) allows us to characterize the transition point as in equation~(\ref{eq:TransitionEnergy}) and such a simple expression is useful as guiding principles to develop tough polymer materials.
In addition, we have elucidated the physical origin of the velocity jump as a dynamic glass transition in the vicinity of the propagating crack tip (see, Figs.~5 and 6).
Our result implies that the discontinuous transition in the crack propagation velocity is a universal phenomenon that could be observed in a broad class of viscoelastic materials.

The present results are useful both from practical and fundamental viewpoints. 
(i) Conventionally the development of new materials tends to be achieved by trials and errors; 
however, the expressions characterizing the marked points on the curve in Fig.~4a are simple enough to remove such trials and errors, 
and pave the way for a more efficient development of tough polymer materials.
(ii) The minimal model proposed in this article is not restricted to the fixed-grip geometry;
we can easily handle other types of crack experiments in the present framework by considering the time dependence of applied strain $\varepsilon$.
For example, tensile and cyclic experiments are treated by setting $\varepsilon(t)=vt$ and $\varepsilon(t)=A\sin(\omega t)$, respectively.
Here, $v$ is the tensile velocity, $A$ is the amplitude, and $\omega$ is the angular frequency.
We will discuss this line of research elsewhere. 
(iii) The present results involve an interesting analogy to conventional phase transitions.
There appear two quantities $\xi_{\Lambda}$ and $\xi_{1}$ associated with the front and rear sides, respectively, that play a role for the order parameter of the velocity jump in a sense that it changes form one characteristic value to the other as a function of an external control parameter (see Supplementary Fig.~S4c).
(iv) Connection to reaction-diffusion systems is an important issue to be explored. 
Equation~(\ref{eq:EoM_Voigt2}) in Methods for Kelvin-Voigt limit ($\lambda \to \infty$) belongs to the class of reaction-diffusion equation,
$\frac{\partial}{\partial t}u=D\frac{\partial^2}{\partial x^2}u+R[u]$,
and the counterpart for arbitrary $\lambda$ forms a generalized class.
Accordingly, the present generalization could enrich physical scenarios in reaction-diffusion systems in different disciplines, e.g., pattern formation in chemical reaction systems and morphogenesis in biology.
In return, crack problems in viscoelastic materials can benefit from the field of reaction-diffusion systems. 
The present crack problem corresponds to a linear reaction term $R[u]\propto u$, and nonlinear extension (e.g., Ramberg-Osgood stress-strain relation) is important for dealing with more practical materials. 
Such an extension could be solved with the aid of the accumulated mathematical knowledge in a well-developed field of reaction-diffusion systems \cite{RDE}.\\

\section*{Methods}

{\bf Derivation of the relation between $w$ and $V$.}

To explain how to derive the exact relation between $w$ and $V$ given in \eq~(\ref{eq:Result-Zener}),
we first consider a more simplified model consisting of Kelvin-Voigt elements illustrated in Fig.~2f.
This simpler model is obtained from the present model in the limit $\lambda \to \infty$.
Although this simpler model does not reproduce the velocity jump (see Fig.~4b and 4e), 
it is useful to understand the mathematical structure of the present model.

In this simpler model, the equation of motion of lattice points in the $y$-direction is given by \eq~(\ref{eq:EoM2}) with \eq~(\ref{eq:Voigt}).
%
Thus, the equations of motion (divided by a constant $\alpha$) are given by
\begin{equation}
\begin{cases}
\displaystyle
0= k\frac{\partial^{2}}{\partial x^{2}}u(x,t)
 + c - lE_{0}u(x,t)-l\eta\frac{\partial}{\partial t}u(x,t) \\
 \displaystyle
0= k\frac{\partial^{2}}{\partial x^{2}}u(x,t)
 + c - L E_{0}u(x,t)-L\eta\frac{\partial}{\partial t}u(x,t),
\label{eq:EoM_Voigt2}
\end{cases}
\end{equation}
for the rear and front sides, respectively.
Here, $k\equiv l^{2}K/\alpha$ and $c$ are independent of position ($x$) and time ($t$).
To seek a solution corresponding to a constant-velocity crack propagation, 
we substitute the form $u(x,t)=f(x-Vt)$ into Eq.~(\ref{eq:EoM_Voigt2})
to obtain linear ordinary differential equations (ODE):
\begin{equation}
\begin{cases}
\displaystyle
0=c-lE_{0}f(x)+lV\eta f'(x)+ kf''(x) \\
\displaystyle
0=c-LE_{0}f(x)+LV\eta f'(x)+ kf''(x),
\label{eq:EoM_Voigt3}
\end{cases}
\end{equation}
for the rear and front sides, respectively.

We can solve equation~(\ref{eq:EoM_Voigt3}) with appropriate boundary conditions at $x=\pm\infty$ and matching conditions for the rear and front solutions at the crack tip (See Supplementary Section~II for the details).
As a result, we find that crack propagates only in the range 
$1/\sqrt{\Lambda}<\tilde{\varepsilon}<1$ 
or equivalently $w_{0}<w<w_{0}\Lambda$, 
and the velocity is exactly given by
\begin{equation}
\frac{V}{V_{0}}
=\frac{\Lambda\tilde{\varepsilon}^{2}-1}
          {\sqrt{\Lambda\tilde{\varepsilon}\left(  1-\tilde{\varepsilon}\right)
                     \left( \Lambda \tilde{\varepsilon}-1\right)}}, 
\label{eq:Result-Voigt}
\end{equation}
with $\tilde{\varepsilon} \equiv\varepsilon/\varepsilon_{c}=\sqrt{w/(w_{0}\Lambda)}$.
Equation~(\ref{eq:Result-Voigt}) for the model consisting of Kelvin-Voigt elements
is the counterpart of \eq~(\ref{eq:Result-Zener}) for the model consisting of Zener elements.
In fact, by taking the limit $\lambda \to \infty$ in \eq~(\ref{eq:Result-Zener}),
we have \eq~(\ref{eq:Result-Voigt}), which does not reproduce the velocity jump (Fig.~4e), unlike \eq~(\ref{eq:Result-Zener}).

We next briefly describe how to generalize the above procedure to the model consisting of Zener elements illustrated in Fig.~2e.
The counterparts of equation~(\ref{eq:EoM_Voigt2}) is expressed as the following set of equation of motion, in which two variables $u$ and $u_2$ are coupled:
\begin{equation}
\begin{cases}
\displaystyle
0= k\frac{\partial^{2}}{\partial x^{2}}u(x,t)
 + c - lE_{0}u(x,t)-l\eta\frac{\partial}{\partial t}u_{2}(x,t) \\
 \displaystyle
0= k\frac{\partial^{2}}{\partial x^{2}}u(x,t)
 + c - LE_{0}u(x,t)-L\eta\frac{\partial}{\partial t}u_{2}(x,t).
\end{cases}
\end{equation}
Here, $E_{1}u_{1}=\eta\frac{\partial}{\partial t}u_{2}$, 
with the elongation of dashpot $u_{2}$ and the total elongation $u=u_{1}+u_{2}$. 
By noting the relation $u=\frac{\eta}{E_{1}}\frac{\partial}{\partial t}u_{2}+u_{2}$, 
the set of equation of motion can be written only in terms of $u_{2}$ by removing the variables $u$ and $u_{1}$.
Introducing the form $u_{2}(x,t)=f_2(x-Vt)$ as before, 
we obtain a third-order linear ODE for $f_2$, which can be solved under the boundary conditions including matching conditions for the rear and front solutions.
As a result, we obtain \eq~(\ref{eq:Result-Zener}) together with \eq~(\ref{eq:CharacteristicEq}),
which is a characteristic equation of the third-order linear ODE
for $f_2$.
We explain the details of the derivation in Supplementary Section~III.\\

{\bf $\Psi_{\mathrm{soft}}$ and $\Psi_{\mathrm{hard}}$ for short and long Zener elements}

We give explicit forms of the parameters $\Psi_{\mathrm{soft}}$ and $\Psi_{\mathrm{hard}}$ used to plot Fig.~5b-c.
By using results obtained in Supplementary Section~III, we have $\Psi_{\mathrm{soft}}$ and $\Psi_{\mathrm{hard}}$ for ``short'' Zener elements as
\begin{equation}
\begin{cases}
\Psi_{\mathrm{soft}}^{\mathrm{short}}
= \frac{\lambda -1}{\lambda}
\frac{\xi_{1}}{\nu}
\left[  \frac{\tilde{\varepsilon}}{1-\tilde{\varepsilon}}   e^{\chi/\xi_1} +1 \right]\\
\Psi_{\mathrm{hard}}^{\mathrm{short}}
= 
\frac{\nu}{(\lambda -1)\xi_{1}}\left[
      \frac{\tilde{\varepsilon}}{1-\tilde{\varepsilon}}
         \cdot
   \frac{(\lambda-1)\xi_{1}+\nu}
          {(\lambda-1)\xi_{1}+\lambda\nu}
      e^{\chi/\xi_1}
      +1
\right]^{-1} \!\!,
\end{cases}
\label{eq:Psi-Short}
\end{equation}
respectively.
Here, $\nu\equiv V/V_0$, $\tilde{\varepsilon}\equiv \varepsilon/\varepsilon_c=\sqrt{w/(w_0 \Lambda)}$,
and
$\chi\equiv x/x_0$ is the distance along the $x$-axis from the crack tip normalized by reference length scale $x_{0}\equiv l\sqrt{\left(1-\frac{1}{N}\right)\frac{\mu}{2E_{0}}}$.
Equations~(\ref{eq:Psi-Short}) together with \eq~(\ref{eq:Result-Zener}) give contour plots in Fig.~5b.

Expressions for the ``long'' Zener elements are different depending on whether the element is located at the front or rear side of the crack tip.
On the front side, we have
\begin{equation}
\begin{cases}
\Psi_{\mathrm{soft}}^{\mathrm{long, front}}
= \frac{\lambda -1}{\lambda}
\frac{\xi_{1}}{\nu}
\left[
      \frac{(\Lambda-1)\tilde{\varepsilon}}{1-\tilde{\varepsilon}}
      e^{\chi/\xi_1}
      -1
\right]\\
\textstyle\Psi_{\mathrm{hard}}^{\mathrm{long, front}}
=\frac{\nu}{(\lambda -1)\xi_{1}}
\left[
      \frac{(\Lambda-1)\tilde{\varepsilon}}{1-\tilde{\varepsilon}}
         \cdot
   \frac{(\lambda-1)\xi_{1}+\nu}
          {(\lambda-1)\xi_{1}+\lambda\nu}
      e^{\chi/\xi_1}
      -1
\right]^{-1}\!.
\end{cases}
\label{eq:Psi-LongFront}
\end{equation}
On the rear side, we have
\begin{equation}
\begin{cases}
\Psi_{\mathrm{soft}}^{\mathrm{long, rear}}
\equiv \frac{\lambda -1}{\lambda}
\frac{\mathcal{E}}{\left|\dot{\mathcal{E}}\right|}\\
\Psi_{\mathrm{hard}}^{\mathrm{long, rear}}
\equiv \frac{1}{\lambda -1}
\frac{\left|\dot{\sigma}\right|}{\sigma},
\end{cases}
\label{eq:Psi-LongRear}
\end{equation}
where
\begin{equation}
\begin{cases}
\mathcal{E}
= C_{0}  
\sum_{i=1}^2 D_{i}
\left[1+\frac{\nu}{(\lambda -1)\xi_{N,i}}\right]
e^{-\chi /\xi_{N,i}} \\
\dot{\mathcal{E}}
=
C_{0} 
\sum_{i=1}^2 D_{i}
\left[
1+\frac{\nu}{(\lambda -1)\xi_{N,i}}
\right]
\frac{\nu}{\xi_{N,i}}
e^{-\chi /\xi_{N,i}}\\
\mathcal{\sigma}
= C_{0}  
\sum_{i=1}^2 D_{i}
\left[1+\frac{\lambda\nu}{(\lambda -1)\xi_{N,i}}\right]
e^{-\chi /\xi_{N,i}} \\
\dot{\mathcal{\sigma}}
=
C_{0} 
\sum_{i=1}^2 D_{i}
\left[
1+\frac{\lambda\nu}{(\lambda -1)\xi_{N,i}}
\right]
\frac{\nu}{\xi_{N,i}}
e^{-\chi /\xi_{N,i}}.
\end{cases}
\label{eq:solutions-long-rear}
\end{equation}
Note that $\chi <0$ on the rear side.
Here, $\xi_{N,1}$, $\xi_{N,2}$, and $\xi_{N}$ with $\xi_{N,1}<\xi_{N,2}<0<\xi_{N}$ are the solutions of the cubic \eq~(\ref{eq:CharacteristicEq}) for $\xi$ 
with $D_{1}= \frac{\gamma_{2}+1}{\gamma_{1}\left( \gamma_{2} - \gamma_{1} \right)}$ and $D_{2}= -\frac{\gamma_{1}+1}{\gamma_{2}\left( \gamma_{2} - \gamma_{1} \right)}$
where $\gamma_{1}\equiv - \xi_{1}/\xi_{N,1}$ and $\gamma_{2}\equiv - \xi_{1}/\xi_{N,2}$.
$C_0\equiv \frac{\varepsilon_c-\varepsilon}{\Lambda-1}
   \cdot
   \frac{(\lambda-1)\xi_{1}}{(\lambda-1)\xi_{1}+\nu}$ is a positive constant.
Equations~(\ref{eq:Psi-LongFront}) and (\ref{eq:Psi-LongRear}) together with \eq~(\ref{eq:Result-Zener}) give contour plots in Fig.~5c.\\

\textit{Note added} ---
After completion of this work, a finite-element-method (FEM) study that numerically reproduces the velocity jump was reported in Ref.~\cite{KuboUmeno2017} with taking into account nonlinear viscoelasticity introducing 30 material parameters.
Unlike their model, since we analyzed a minimal model based on linear viscoelasticity with three material parameters ($E_0$, $E_\infty$, and $\eta$), we can solve exactly and probide a clear physical understanding of the phenomenon.

\begin{acknowledgements}
The authors thank Katsuhiko~Tsunoda for providing us with experimental
data that we reproduced in Fig.~1e. The authors thank Katsuhiko~Tsunoda,
Yoshihiko~Morishita, Kohzo~Ito, Kenji~Urayama, Hiroya~Kodama, Atsushi~Kubo,
Yoshitaka Umeno, Jian Ping Gong, Atsushi Takahara, and Yuko Aoyanagi for fruitful discussions. 
N.S. thanks Hiroki~Fukagawa, Tetsuo~Hatsuda, Atsushi~Ikeda, Kyogo~Kawaguchi, Takashi~Mori, Akira~Shimizu, and Hal~Tasaki for useful comments,
and is supported by JSPS KAKENHI Grant Number~15K17725. 
This research was partly supported by Grant-in-Aid for
Scientific Research (A) (No.~24244066) of JSPS, Japan, and by ImPACT Program
of Council for Science, Technology and Innovation (Cabinet Office, Government
of Japan).
\end{acknowledgements}

\widetext
\pagebreak
\widetext
\clearpage

\widetext 

\begin{center}
\textbf{\large Supplementary Information for:\\
Exactly solvable model for a velocity jump \\
observed in crack propagation in viscoelastic  solids}
\end{center}
\setcounter{section}{0}
\setcounter{equation}{0}
\setcounter{figure}{0}
\setcounter{table}{0}
\setcounter{page}{1}
\makeatletter
\renewcommand{\theequation}{\thesection.\arabic{equation}}
\renewcommand{\thefigure}{S\arabic{figure}}
\renewcommand{\bibnumfmt}[1]{[S#1]}
\renewcommand{\citenumfont}[1]{S#1}
 \@addtoreset{equation}{section}

In this note, we explain the details of the formulation of the model and the derivation of the analytical solutions given in the main text. 
This note is organized as follows. 
In Sec.~I, we describe a lattice model for the fixed-grip crack propagation in viscoelastic sheets with incorporating Kelvin-Voigt elements for the interaction between the lattice points. 
In Sec.~II, we construct a model for which an analytical solution is available for a constant-velocity crack propagation, 
by simplifying the ordinary lattice model introduced in Sec. I. 
For the model consisting of Kelvin-Voigt elements, 
we derive an analytical expression for the crack-propagation velocity as a function of the initially applied energy density
and show that the model does not exhibit the velocity jump. 
In Sec.~III, we replace the Kelvin-Voigt interaction with the Zener interaction in the model.
For the model consisting of Zener elements, 
we derive an analytical expression for the crack-propagation velocity as a function of the initially applied energy density
and reveal that the model exhibits the velocity jump. 
In Sec.~IV, from the analytical solution, we derive the existence condition of the velocity jump, 
together with simple relations useful for controlling crack propagation in developing tough materials.

\begin{figure}[b!]
 \centering
\includegraphics[keepaspectratio, scale=1.0]{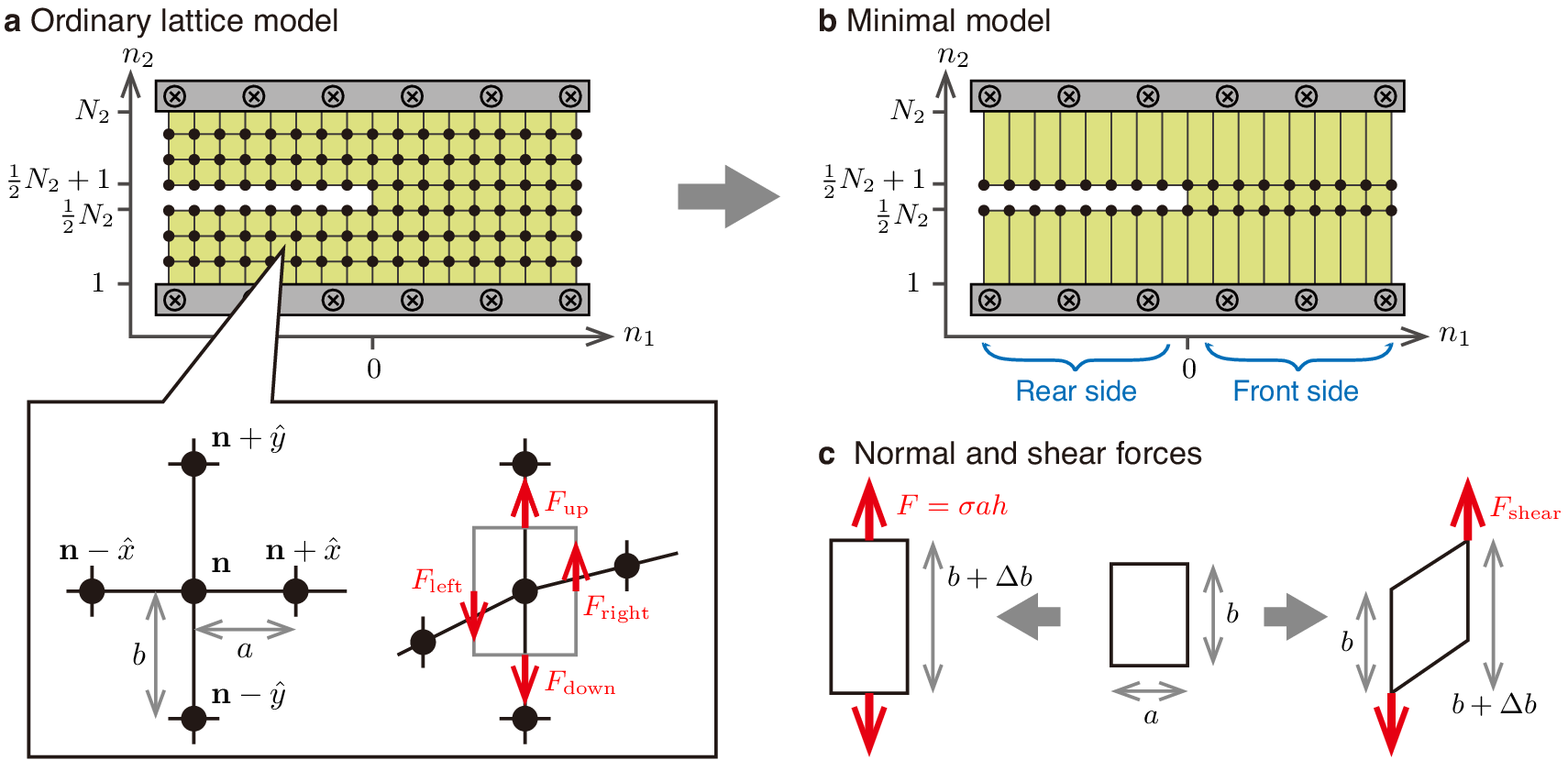}
 \caption{
\textbf{Ordinary and minimal lattice models for the fixed-grip crack propagation.}
\textbf{a}, Lattice model with a rectangular unit cell under zero strain. 
The lattice point is specified by two integers $\mathbf{n}=(n_{1},n_{2})$. 
The top and bottom boundaries are located at $n_{2}=1$ and $n_{2}=N_{2}$, respectively, and the two surfaces of the line crack are located at $n_{2}=N_{2}/2$ and
$n_{2}=N_{2}/2+1$. 
In the box, magnified views on a lattice point and its nearest neighbor points with and without deformation are shown together with the indices of the points, the lattice constants $a$ and $b$, and associated forces.
\textbf{b}, Coarse-grained semi-lattice model obtained by removing lattice points except for the
ones located at $n_{2}=N_{2}/2$ and $n_{2}=N_{2}/2+1$. 
\textbf{c}, Elementary deformation modes of a unit cell in the rectangular lattice with lattice spacings $a$ and $b$.}
\label{fig:sup:Model}
\end{figure}

\section{Lattice model consisting of Kelvin-Voigt elements}

We construct a lattice model for crack propagation in a viscoelastic sheet,
which satisfies the following conditions:
\begin{itemize}
\item The viscoelastic sheet is always on the $x$-$y$ plane. Under zero strain,
the height (in the $y$-direction) and thickness (in the $z$-direction) of the
sheet are $L$ and $h$, respectively, whereas the width (in the $x$-direction)
is much larger than $L$.
\item The fixed strain $\varepsilon$ ($\equiv\Delta L/L\geq0$) is applied in
the $y$-direction.
\item A line crack propagates in the positive $x$-direction.
\end{itemize}

We consider a two-dimensional model on an $N_{1}\times N_{2}$ rectangular lattice where $N_{1}$ is (countably) infinite and $N_{2}$ is even. 
The lattice points are labeled by the index $\mathbf{n}=(n_{1},n_{2})$ 
where $n_{1}\in \mathbb{Z}\equiv\{\dots,-1,0,1,\dots\}$ and $n_{2}\in\{1,2,\dots,N_{2}\}$. 
At $\varepsilon=0$, the lattice constant in the $x$- and $y$-direction are $a$ and $b$, respectively. 
In other words, under zero strain, the height and width of the sheet are $aN_{1}(=\infty)$ and $bN_{2}(=L)$, respectively. 
The position of the lattice point labeled by the index $\mathbf{n}$ is given by
$\mathbf{r}_{0}(\mathbf{n})=(an_{1},bn_{2}-L/2)$. 
When the top and bottom edges of the sheet are clamped and stretched in the $y$-direction
($\varepsilon>0$), we set the coordinate of the points on the edges to
$\mathbf{r}_{\varepsilon}(n_{1},1)=(an_{1},-L(1+\varepsilon)/2)$ and
$\mathbf{r}_{\varepsilon}(n_{1},N_{2})=(an_{1},L(1+\varepsilon)/2)$. 
The displacement of the lattice point $\mathbf{n}$ is given by 
$\mathbf{r}_{\varepsilon}(\mathbf{n})-\mathbf{r}_{0}(\mathbf{n})\equiv\mathbf{u}(\mathbf{n})=(u_{1}(\mathbf{n}),u_{2}(\mathbf{n}))$.
Note that, in the main text, we use $u_{i}$ to denote the $y$-coordinate.

In the model, every lattice point interacts with the nearest neighbor points (at
most four points) by a minimal coupling.
The tensile and shear stresses are given as $F_{\mathrm{tensile}}/(ah)=E\Delta b/b$ and
$F_{\mathrm{shear}}/(bh)=G\Delta b/a$, respectively, for the deformation characterized by $\Delta b$ 
(See, Fig.~\ref{fig:sup:Model} for the case in which the unit cell deforms in the $y$-direction). 
Here, we have introduced Young's modulus $E$ and the shear modulus $G$. The
elastic energy of the sheet with the minimal coupling is then given by
\begin{equation}
U[u_{\mu}(\mathbf{n})] 
=h\sum_{n_{1}=-\infty}^{\infty}
\sum_{n_{2}=1}^{N_{2}}
\sum_{\mu=1}^{2}
\sum_{\nu=1}^{2} 
\frac{1}{2}C_{\mu\nu}
\left[  u_{\mu}(\mathbf{n}+\hat{\nu})-u_{\mu}(\mathbf{n})\right]^{2},
\label{eq:sup:potential}
\end{equation}
where $C_{11}=C_{22}= Ea/b$
and $C_{12}=C_{21}= Gb/a$.
Here,
$\hat{1}\equiv\hat{x}\equiv(1,0)$ and $\hat{2}\equiv\hat{y}\equiv(0,1)$,
$\delta_{\mu\nu}$ is the Kronecker delta, and the summation extends over all the bonds on the sheet. 
We note that Poisson's ratio of this system is zero, i.e., $G=E/2$ 
because Eq.~(\ref{eq:sup:potential}) does not contain the coupling between the displacements in the $x$- and $y$-directions.
Moreover, assuming that the displacements in the $x$-direction $u_{1}(\mathbf{n})$ for arbitrary $\mathbf{n}$ are zero at the initial time,
forces in the $x$-direction are always zero, and thus, every lattice point does not move in the $x$-direction.
Then, in the following, we consider the motions and forces only in the $y$-direction. 


To construct the equation of motion for the system, 
we consider tensile and shear forces, together with viscous forces.
As illustrated in Fig.~\ref{fig:sup:Model}, the tensile and shear forces acting on the lattice point
$\mathbf{n}$ in the $y$-direction are given, respectively, as follows:
\begin{equation}
F_{\mathrm{up}}-F_{\mathrm{down}}
=Eah\left[  \frac{u_{2}(\mathbf{n}+\hat{y})-u_{2}(\mathbf{n})}{b}-\frac{u_{2}(\mathbf{n})-u_{2}(\mathbf{n}-\hat{y})}{b}\right]
=Eabh\frac{\Delta^{2}}{\Delta y^{2}}u_{2}(\mathbf{n}),
\label{eq:sup-lattice-updownF}
\end{equation}
\begin{equation}
F_{\mathrm{right}}-F_{\mathrm{left}}
=Gbh\left[  \frac{u_{2}(\mathbf{n}+\hat{x})-u_{2}(\mathbf{n})}{a}-\frac{u_{2}(\mathbf{n})-u_{2}(\mathbf{n}-\hat{x})}{a}\right]
=Gabh\frac{\Delta^{2}}{\Delta x^{2}}u_{2}(\mathbf{n}).
\label{eq:sup-lattice-rightleftF}
\end{equation}
Here, we have introduced the notations, 
$\frac{\Delta^{2}}{\Delta x^{2}}f(\mathbf{n})
\equiv\left[  f(\mathbf{n}+\hat{x})-2f(\mathbf{n})+f(\mathbf{n}-\hat{x})\right]  /a^{2}$
and $\frac{\Delta^{2}}{\Delta y^{2}}f(\mathbf{n})
\equiv\left[  f(\mathbf{n}+\hat{y})-2f(\mathbf{n})+f(\mathbf{n}-\hat{y})\right]  /b^{2}$, 
which correspond to the second-order partial
derivatives in the continuum limits, $a\rightarrow0$ and $b\rightarrow0$,
respectively. 
When a Kelvin-Voigt element is employed for the interaction in
the $y$-direction, the following viscous term should be added:
\begin{equation}
F_{\mathrm{up}}^{(\eta)}-F_{\mathrm{down}}^{(\eta)}
=\eta ah\left[  \frac{\frac{\partial}{\partial t}u_{2}(\mathbf{n}+\hat{y})
-\frac{\partial}{\partial t}u_{2}(\mathbf{n})}{b}-\frac{\frac{\partial}{\partial t}u_{2}(\mathbf{n})-\frac{\partial}{\partial t}u_{2}(\mathbf{n}-\hat{y})}{b}\right]
 =\eta abh\frac{\Delta^{2}}{\Delta y^{2}}\frac{\partial}{\partial t}u_{2}(\mathbf{n}).
\end{equation}
In this way, the equation of motion of the lattice point $\mathbf{n}$ in the
$y$-direction is given by
\begin{equation}
\begin{split}
m\frac{\partial^{2}}{\partial t^{2}}u_{2}(\mathbf{n})
&=F_{\mathrm{right}}-F_{\mathrm{left}}+F_{\mathrm{up}}-F_{\mathrm{down}} 
+ F_{\mathrm{up}}^{(\eta)}-F_{\mathrm{down}}^{(\eta)}
\\
&=abh\left[  G\frac{\Delta^{2}}{\Delta x^{2}}u_{2}
(\mathbf{n})+E\frac{\Delta^{2}}{\Delta y^{2}}u_{2}(\mathbf{n})+\eta
\frac{\Delta^{2}}{\Delta y^{2}}\frac{\partial}{\partial t}u_{2}(\mathbf{n}%
)\right]  .
\end{split}
\end{equation}
Assuming that we can neglect the inertial term $m\frac{\partial^{2}}{\partial t^{2}}u_{2}(\mathbf{n})$ (the overdamped limit), 
we obtain the equation of motion per unit volume in the following form:
\begin{equation}
0=G\frac{\Delta^{2}}{\Delta x^{2}}u_{2}(\mathbf{n})+E\frac{\Delta^{2}}{\Delta
y^{2}}u_{2}(\mathbf{n})+\eta\frac{\Delta^{2}}{\Delta y^{2}}\frac{\partial
}{\partial t}u_{2}(\mathbf{n}).\label{eq:sup-lattice-model}%
\end{equation}

\section{Minimal model consisting of Kelvin-Voigt elements}
\label{sec:sup:KV}

\subsection{Construction of a minimal model}

In this section, 
we construct a minimal model for crack propagation in viscoelastic sheets as shown in Fig.~\ref{fig:sup:Model}b,
by ignoring all the lattice points except for the lattice points at
$n_{2}=N_{2}/2$ and $n_{2}=N_{2}/2+1$.
The original set of variables $\{u_{2}(\mathbf{n})\}$ is now represented by a much smaller set, 
$\{u_{2}(i,N_{2}/2)\}\cup\{u_{2}(i,N_{2}/2+1)\}$ for $i\in\mathbb{Z}\equiv\{\dots,-1,0,1,\dots\}$. 
For simplicity, we assume that the sheet is always symmetric about the $x$ axis,
and thus the lattice points on the upper side, $u_{i}\equiv u_{2}(i,N_{2}+1)$ for $i\in\mathbb{Z}$,
completely describe the dynamics of the present model.

\begin{figure}[b!]
\centering
\includegraphics[keepaspectratio, scale=1]{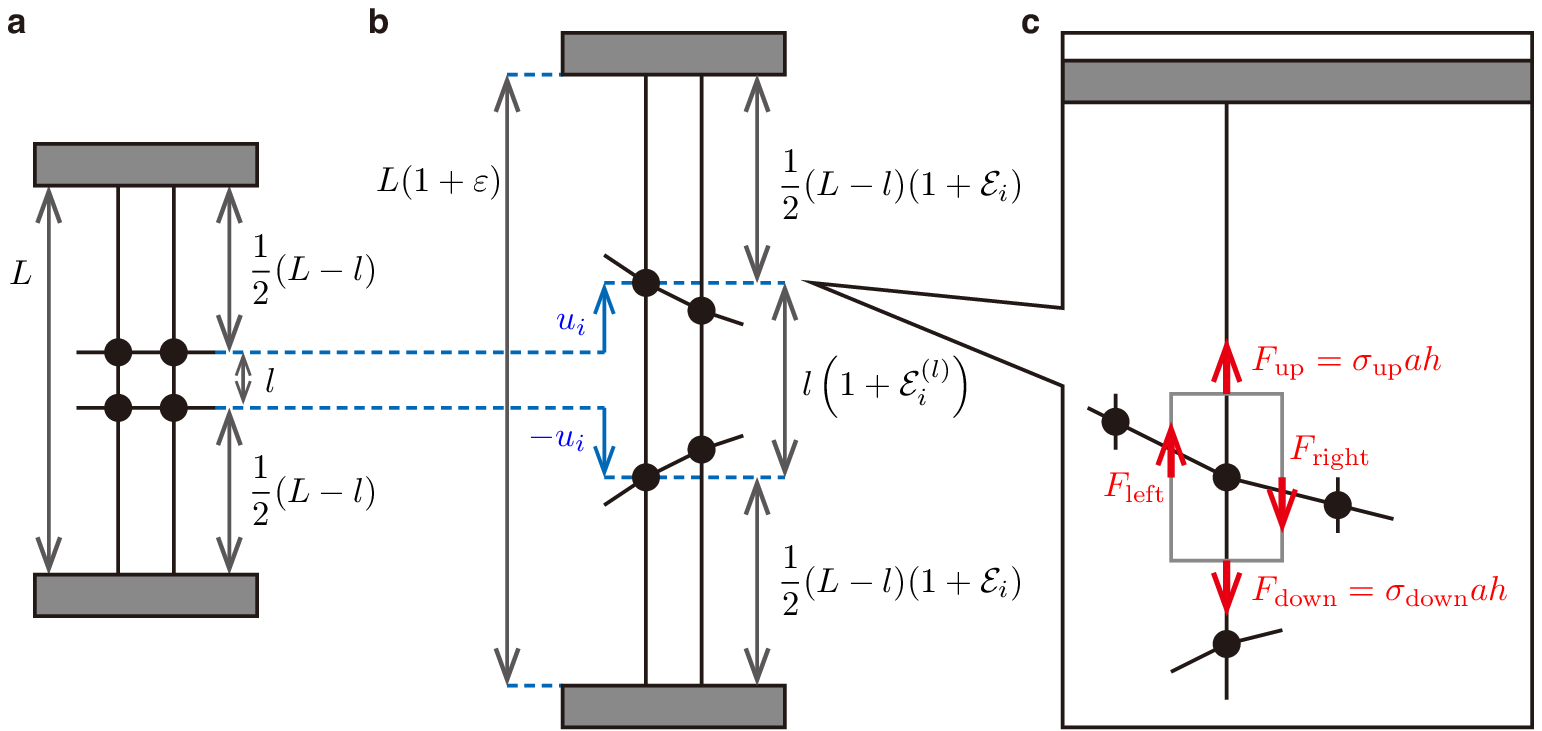}  
\caption{
\textbf{Three types of strain
($\varepsilon$, $\mathcal{E}_{i}(t)$, and $\mathcal{E}_{i}^{(l)}(t)$)
and forces.}
\textbf{a,} The proposed minimal lattice model under zero strain. 
We call the springs of length $(L-l)/2$ and those of length $l$ the long and short springs, respectively. 
\textbf{b,} The model under the finite strain $\varepsilon$. 
\textbf{c,} Forces acting on a lattice point in the model.
}
\label{fig:sup:ThreeKindOfStrain}
\end{figure}

In the following discussion, it is important to distinguish three types of
strain in the $y$-direction (see Fig.~\ref{fig:sup:ThreeKindOfStrain}a; 
here and hereafter, we set $l \equiv b$ and use $l$ instead of $b$):


\begin{enumerate}
\item[(i)] $\varepsilon$: the global fixed strain;
\item[(ii)] $\mathcal{E}_{i}(t)$: the strain of the $i$-th upper (or lower) ``long spring," i.e.,
the $i$-th spring of natural length $(L-l)/2$ directly connected to
the top (or bottom) boundary (the strains of the upper and lower long springs are the same so that
we consider only the upper ones and ``long springs" indicate the upper ones);
\item[(iii)] $\mathcal{E}_{i}^{(l)}(t)\equiv\lbrack u_{2}(i,N_{2}+1)-u_{2}(i,N_{2})]/l=2u_{i}/l$:
 the strain of the $i$-th ``short spring," i.e., the
$i$-th spring of natural length $l$ located at the center in the
$y$-direction (provided that the $i$-th short spring exists).
\end{enumerate}
Here, $\varepsilon$ is a constant, 
but $\mathcal{E}_{i}(t)$ and $\mathcal{E}_{i}^{(l)}(t)$ depend on time $t$.
We mainly use $\mathcal{E}_{i}(t)$ to describe the dynamics of crack propagation
because $\mathcal{E}_{i}^{(l)}(t)$ is expressed by $\mathcal{E}_{i}(t)$ by the following relation (see Fig.~\ref{fig:sup:ThreeKindOfStrain}b):
\begin{equation}
L\varepsilon=(L-l)\mathcal{E}_{i}(t)+l\mathcal{E}_{i}^{(l)}(t).
\label{eq:sup:epsilonepsilonepsilon}%
\end{equation}
Note that $\mathcal{E}_{i}(t)$ is much smaller than $\mathcal{E}_{i}^{(l)}(t)$ provided that the $i$-th short spring exists because ``the spring constant'' of long springs is much smaller than that of the short springs (the corresponding Young's moduli are the same for the short and long springs).
To realize crack propagation, we remove the $i$-th short spring when 
$\mathcal{E}_{i}^{(l)}(t)\geq\varepsilon_{c}$. 
At places far from the crack tip, 
$\mathcal{E}_{i}(t)$ approaches a constant value and the value depends on 
whether the $i$-th long spring is located on the front (i.e., right) or rear (i.e., left)
side of the tip of a crack propagating in the positive
$x$-direction (see Fig.~\ref{fig:sup:Model}b):
\begin{equation}%
\begin{cases}
\displaystyle\lim_{i\rightarrow-\infty}\mathcal{E}_{i}(t)=0 
& \text{rear side}\\
\displaystyle\lim_{i\rightarrow+\infty}\mathcal{E}_{i}(t)=\lim_{i\rightarrow
+\infty}\mathcal{E}_{i}^{(l)}(t)=\varepsilon .
\qquad 
& \text{front side}%
\end{cases}
\label{eq:sup:epsilonFarFromTip}
\end{equation}

We construct the equation of motion in the $y$-direction for the $i$-th lattice point characterized by $u_{i}(t)$ in the minimal model.
First, we consider tensile stresses on the basis of Eq.~(\ref{eq:sup-lattice-model});
as illustrated in Fig.~\ref{fig:sup:ThreeKindOfStrain}c,
considering the elongational deformation in the $y$-direction of the $i$-th
long and short springs, we obtain the tensile stresses acting on the $i$-th
lattice point in the following form:
\begin{equation}
\begin{cases}
\displaystyle\frac{F_{\mathrm{up},i}(t)}{ah}=E_{0}\mathcal{E}_{i}(t)\\
\displaystyle\frac{F_{\mathrm{down},i}(t)}{ah}=E_{0}\mathcal{E}_{i}^{(l)}(t),
\end{cases}
\end{equation}
from which, we have
$F_{\mathrm{up},i}(t)-F_{\mathrm{down},i}(t)
=E_{0}\left(  \mathcal{E}_{i}(t)-\varepsilon\right)ahL/l$
by virtue of Eq.~(\ref{eq:sup:epsilonepsilonepsilon}). 
Here, we display the subscript $0$ for the Young's modulus, $E_{0}$,
whereas we have omitted the subscript in Sec.~I, i.e., $E \equiv E_{0}$.
This is because we should distinguish the two springs $E_{0}$ and $E_{1}$ in a Zener element, which is a generalization of a Kelvin-Voigt element and will be considered in Secs.~III and IV.
Second, we consider shear stresses; 
as illustrated in Fig.~\ref{fig:sup:ThreeKindOfStrain}c, 
considering the shear deformation in the $y$-direction, 
we obtain 
$F_{\mathrm{right},i}-F_{\mathrm{left},i}=alh\mu\frac{\Delta^{2}}{\Delta x^{2}}u_{i}(t)$. 
Note that shear modulus $G$ for the ordinary lattice model (Fig.~\ref{fig:sup:Model}a, discussed in Sec.~I)
is replaced by the ``effective'' shear modulus $\mu$ for the minimal model (Fig.~\ref{fig:sup:Model}b).
This is because the shear modulus for the minimal model is considered to effectively represent all the forces acting on the decimated points from the nearest neighbor points located in the $x$-directions, 
with a spirit similar to the one employed in renormalization \cite{S_Cardy}
in statistical physics.
Third, combining the tensile and shear stresses together with the viscous terms, 
we obtain the equations of motion in the $y$-direction for the $i$-th lattice point of mass $m$, 
by noting that $F_{\mathrm{down},i}(t)$ and $F_{\mathrm{down},i}^{(\eta)}(t)$ are missing on the rear side of the crack tip:
\begin{equation}%
\begin{cases}
\displaystyle 
m\frac{\partial^{2}}{\partial t^{2}}u_{i}(t)
=F_{\mathrm{up},i}+ F_{\mathrm{up},i}^{(\eta)}+F_{\mathrm{right},i}-F_{\mathrm{left},i}
=ah\left[E_{0}\mathcal{E}_{i}(t)
+\eta\frac{\partial}{\partial t}\mathcal{E}_{i}(t)\right]
+alh\mu\frac{\Delta^{2}}{\Delta x^{2}}u_{i}(t) 
& \text{rear side}\\
\begin{split}
\displaystyle m\frac{\partial^{2}}{\partial t^{2}}u_{i}(t)
&=F_{\mathrm{up},i}+F_{\mathrm{up},i}^{(\eta)}-F_{\mathrm{down},i}-F_{\mathrm{down},i}^{(\eta)}
+F_{\mathrm{right},i}-F_{\mathrm{left},i}
\\
&=ah\frac{L}{l}\left[E_{0}\left(  \mathcal{E}_{i}(t)-\varepsilon\right)
+\eta\frac{\partial}{\partial t}\mathcal{E}_{i}(t)\right]
 +alh\mu\frac{\Delta^{2}}{\Delta x^{2}}u_{i}(t)\quad 
 \end{split}
& \text{front side.}
\end{cases}
\end{equation}
Since Eq.~(\ref{eq:sup:epsilonepsilonepsilon}) can be expressed as
$\mathcal{E}_{i}(t)=[L\varepsilon-2u_{i}(t)]/(L-l)$, we can rewrite the above
equations of motion in terms of $\mathcal{E}_{i}(t)$ by removing the
dynamic variable $u_{i}(t)$. 
Finally, assuming that the inertial term is
negligible (the overdamped limit), we obtain the equations of motion for the
field $\mathcal{E}_{i}(t)$:
\begin{equation}
\begin{cases}
\displaystyle
0
=E_{0}\mathcal{E}_{i}(t)
+\eta\frac{\partial}{\partial t}\mathcal{E}_{i}(t)
-\frac{1}{2}l(L-l)\mu\frac{\Delta^{2}}{\Delta x^{2}}
\mathcal{E}_{i}(t) & \text{rear side}\\
& \\
\displaystyle
0=\frac{L}{l}E_{0}\left(  \mathcal{E}_{i}(t)-\varepsilon\right)
+\frac{L}{l}\eta\frac{\partial}{\partial t}\mathcal{E}_{i}(t)
-\frac{1}{2}l(L-l)\mu\frac{\Delta^{2}}{\Delta x^{2}}\mathcal{E}_{i}(t)\qquad &
\text{front side.}
\end{cases}
\label{eq:sup:EoSperAH}
\end{equation}

We have three comments on the equations of motion~(\ref{eq:sup:EoSperAH}).
(i) When we neglect shear force ($\mu =0$),
the two expressions on the front and rear sides reduce to the same equation of motion
(except for the equilibrium position), which describes the Kelvin-Voigt dynamics.
(ii) The ``second-derivative'' term
$E_{0}\frac{\Delta^{2}}{\Delta y^{2}}u_{2}(\mathbf{n})$ in Eq.~(\ref{eq:sup-lattice-model}) has been replaced by the strain-field term, 
$E_{0}\mathcal{E}_{i}(t)$ (rear side) 
or $\frac{L}{l}E_{0}\left(  \mathcal{E}_{i}(t)-\varepsilon\right)  $ (front side) in Eq.~(\ref{eq:sup:EoSperAH}) with proportional constants.
(iii) The net strains of long springs on the rear and front sides are different
and given by $\mathcal{E}_{i}(t)$ and $\mathcal{E}_{i}(t)-\varepsilon$, respectively,
because $\displaystyle \lim_{i\to -\infty}\mathcal{E}_{i}(t)=0$ and 
$\displaystyle \lim_{i\to -\infty}\mathcal{E}_{i}(t)=\varepsilon$
on the rear and front sides, respectively (see Eq.~(\ref{eq:sup:epsilonFarFromTip})).

To derive the equations to be solved analytically, 
we take the continuum limit of Eq.~(\ref{eq:sup:EoSperAH}) in the $x$-direction, 
$a\rightarrow0$.
In this limit, the finite difference $\frac{\Delta^{2}}{\Delta x^{2}}$ 
is replaced by the derivative $\frac{\partial^{2}}{\partial x^{2}}$ 
and the discrete strain field $\mathcal{E}_{i}(t)$ 
by the continuum strain field $\mathcal{E}\equiv\mathcal{E}(\tau,\chi)$
where we have introduced the dimensionless parameters, $\tau=t/t_{0}$, $\chi=x/x_{0}$
with
\begin{equation}
t_{0}\equiv\frac{\eta}{E_{0}}
\quad\text{and}\quad
x_{0}\equiv l\sqrt{\left(1-\frac{l}{L}\right)\frac{\mu}{2E_{0}}}.
\end{equation}
In this way, the above discrete version of the equations motion~(\ref{eq:sup:EoSperAH}) is replaced by the following continuum equations:
\begin{equation}
\begin{cases}
\displaystyle
0=\mathcal{E}+\dot{\mathcal{E}}-\Lambda\mathcal{E}^{\prime\prime}
& \text{rear side}\\
\displaystyle
0=\left(  \mathcal{E}-\varepsilon\right)  +\dot{\mathcal{E}}-\mathcal{E}^{\prime\prime}. 
\qquad & \text{front side}
\end{cases}
\label{eq:sup:EoM-Voigt-E}
\end{equation}
Here, we have introduced the notations, $\dot{\mathcal{E}}\equiv\frac
{\partial}{\partial\tau}\mathcal{E}(\tau,\chi)$,
$\mathcal{E}^{\prime\prime}\equiv\frac{\partial^{2}}{\partial\chi^{2}}\mathcal{E}(\tau,\chi)$,
and
\begin{equation}
\Lambda\equiv \frac{L}{l}.
\end{equation}

\subsection{Derivation of an exact solution of crack propagation with a constant velocity}

In this subsection, we solve Eq.~(\ref{eq:sup:EoM-Voigt-E}) 
for the static case ($V=0$) 
and 
for the dynamic case in which a crack propagates with a constant velocity
($0<V<\infty$), by seeking the solution of the form $\mathcal{E}(\tau,\chi)=f(\chi-\nu\tau)$. 
Here, we have introduced the dimensionless velocity of crack propagation,
\begin{equation}
\nu=\frac{V}{V_{0}},
\end{equation}
with $V_{0}$ defined by
\begin{equation}
V_{0}\equiv\frac{x_{0}}{t_{0}}=\frac{l}{\eta}\sqrt{\left(1-\frac{l}{L}\right)\frac{E_{0}\mu}{2}}.
\label{eq:sup:def:V0}
\end{equation}
Substituting $f(\chi-\nu\tau)$ into Eq.~(\ref{eq:sup:EoM-Voigt-E}), we have
second-order linear ordinary differential equations:
\begin{equation}%
\begin{cases}
\displaystyle0=f(\chi)-\nu f^{\prime}(\chi)-\Lambda f^{\prime\prime}(\chi)\qquad 
& \text{for}\,\,\, \chi<0\quad\text{({rear side)}}\\
\displaystyle0=f(\chi)-\varepsilon-\nu f^{\prime}(\chi)-f^{\prime\prime}(\chi) 
& \text{for}\,\,\, 0 \leq \chi \quad\text{{(front side)}}.
\end{cases}
\label{eq:sup:eom-KVoigt-steady}
\end{equation}
Here, since the width (in the $x$-direction) of the sheet is large enough, 
no generality is lost by setting the position of crack tip to $\chi=0$ with $\tau=0$: 
the crack exists in the region $\chi<0$ and is absent in the region $\chi\geq 0$.

We give the boundary conditions for the differential equations~(\ref{eq:sup:eom-KVoigt-steady}) as follows: 
the conditions at remote edges, (see Eq.~(\ref{eq:sup:epsilonFarFromTip}))
\begin{equation}%
\begin{cases}
\displaystyle
f(-\infty)=0 
\qquad 
& \text{rear side}\\
\displaystyle
f(+\infty)=\varepsilon 
& \text{front side}
\end{cases}
\label{eq:sup:bc-KVoigt-steady1}
\end{equation}
and the matching conditions at the crack tip for the strain field $\mathcal{E}$,
\begin{equation}%
\begin{cases}
\displaystyle
f(-0)=f(+0)\equiv f(0) \\
\displaystyle
f^{\prime}(-0)=f^{\prime}(+0).
\end{cases}
\label{eq:sup:bc-KVoigt-steady2}
\end{equation}
As previously mentioned,
we remove the short spring (located in the center in the $y$-direction with natural length $l$, see Fig.~\ref{fig:sup:ThreeKindOfStrain}) 
when $\mathcal{E}^{(l)}(t)\geq\varepsilon_{c}$. 
We rewrite this inequality as
\begin{equation}
\mathcal{E}(\tau,\chi)=f(\chi-\nu\tau)
\leq f_{c}\equiv\frac{\Lambda\varepsilon-\varepsilon_{c}}{\Lambda-1},
\end{equation}
by use of Eq.~(\ref{eq:sup:epsilonepsilonepsilon}). 
Therefore, $f(0)$ satisfies 
(i) $f(0)>f_{c}$ for $\nu=0$ and 
(ii) $f(0)=f_{c}$ for $0<\nu<\infty$.

We solve the ordinary differential equation~(\ref{eq:sup:eom-KVoigt-steady})
under the boundary conditions~(\ref{eq:sup:bc-KVoigt-steady1}) and (\ref{eq:sup:bc-KVoigt-steady2}). 
First, we solve the differential equation for $\chi<0$. 
Substituting the form $f(\chi)=Ce^{-\chi/\xi}$ 
into Eq.~(\ref{eq:sup:eom-KVoigt-steady}), 
we obtain the characteristic equation (quadratic equation for $\xi$), 
\begin{equation}
g_{\Lambda}(\xi)=\xi^{2}+\nu\xi-\Lambda=0, 
\label{eq:sup:def:gLambda-KVoigt}
\end{equation}
and the solutions,
\begin{equation}
\xi_{\Lambda,\pm}=\left(-\nu\pm\sqrt{\nu^{2}+4\Lambda}\right)/2.
\label{eq:sup:def:xiLambda-KVoigt}
\end{equation}
From the boundary conditions~(\ref{eq:sup:bc-KVoigt-steady1}) and (\ref{eq:sup:bc-KVoigt-steady2}), 
only the solution $\xi_{\Lambda,-}$ is relevant:
$f(\chi)=f(0)\exp(-\chi/\left\vert\xi_{\Lambda,-}\right\vert )$. 
Second, we solve the differential equation for $\chi\geq0$. 
Substituting the form $f(\chi)-\varepsilon=Ce^{-\chi/\xi}$ into Eq.~(\ref{eq:sup:eom-KVoigt-steady}), 
we obtain the characteristic equation,
$g_{1}(\xi)=\xi^{2}+\nu\xi-1=0$, 
and the solutions, 
$\xi_{1,\pm}=\left(-\nu\pm\sqrt{\nu^{2}+4}\right)/2$. 
From the boundary conditions~(\ref{eq:sup:bc-KVoigt-steady1}) and (\ref{eq:sup:bc-KVoigt-steady2}), 
only the solution $\xi_{1,+}$ is relevant:
$f(\chi)=\varepsilon-\left[\varepsilon-f(0)\right]\exp(-\chi/\xi_{1,+})$.
Finally, we rewrite the matching condition, 
$f^{\prime}(-0)=f^{\prime}(+0)$ as
\begin{equation}
\frac{f(0)}{\left\vert \xi_{\Lambda,-}\right\vert }
=\frac{\varepsilon-f(0)}{\xi_{1,+}}.\label{eq:sup:BCofVoigt}%
\end{equation}
In the following, we determine $f(\chi)$ from Eq.~(\ref{eq:sup:BCofVoigt}) in the cases of $\nu=0$ and $0<\nu<\infty$.

In the static case ($\nu=0$),
we can rewrite Eq.~(\ref{eq:sup:BCofVoigt}) as
$f(0)=\frac{\sqrt{\Lambda}}{\sqrt{\Lambda}+1}\varepsilon$,
by using $\left\vert \xi_{\Lambda,-}\right\vert=\sqrt{\Lambda}$ and $\xi_{1,+}=1$
from Eq.~(\ref{eq:sup:def:xiLambda-KVoigt}).
This gives the following solution:
\begin{equation}
\mathcal{E}(\tau,\chi)
=f(\chi)
=
\begin{cases}
\displaystyle\frac{\varepsilon\sqrt{\Lambda}}{\sqrt{\Lambda}+1}e^{\chi
/\sqrt{\Lambda}}\qquad & \text{{for}}\,\,\chi<0\quad\text{({rear side)}}\\
& \\
\displaystyle\varepsilon-\frac{\varepsilon}{\sqrt{\Lambda}+1}e^{-\chi} &
\text{{for}\thinspace}\,\,0\leq\chi\quad\text{{(front side)}}%
\end{cases}
\end{equation}
For $\nu=0$, the inequality $f(0)>f_{c}$ yields
\begin{equation}
\tilde{\varepsilon}
\equiv\frac{\varepsilon}{\varepsilon_{c}}
<\frac{1}{\sqrt{\Lambda}}.
\label{eq:sup:result_of_stationary_solution}
\end{equation}
Note that Eq.~(\ref{eq:sup:result_of_stationary_solution}) implies that $\tilde{\varepsilon}>1/\sqrt{\Lambda}$
in the case of crack propagation, $\nu>0$.

In the case of crack propagation with a constant velocity $\nu$ ($0<\nu<\infty$),
since $f(0)=f_{c}$,
the exact relationship between $\nu$ and $\tilde{\varepsilon}$ is obtained from Eq.~(\ref{eq:sup:BCofVoigt}) as
\begin{equation}
\nu\equiv\frac{V}{V_{0}}=\frac{\Lambda\tilde{\varepsilon}^{2}-1}{\sqrt
{\tilde{\varepsilon}(1-\tilde{\varepsilon})(\Lambda\tilde{\varepsilon}-1)}},
\label{eq:sup:nuepsilon_for_KV}
\end{equation}
or equivalently
\begin{equation}
\tilde{\varepsilon}
=
\frac
{\sqrt{\nu^2 + 4\Lambda} + \sqrt{\nu^2 + 4}}
{\sqrt{\nu^2 + 4\Lambda} + \Lambda\sqrt{\nu^2 + 4}
-(\Lambda -1)\nu}.
\label{eq:sup:epsilonnu_for_KV}
\end{equation}
The expression~(\ref{eq:sup:nuepsilon_for_KV}) does not exhibit the velocity jump in the physically relevant range of $\tilde{\varepsilon}$, $1/\sqrt{\Lambda}<\tilde{\varepsilon}<1$.
By using Eq.~(\ref{eq:sup:nuepsilon_for_KV}), we obtain the dynamics of the strain distribution:
\begin{equation}
\mathcal{E}(\tau,\chi)=f(\chi-\nu\tau)=%
\begin{cases}
\displaystyle\frac{\Lambda\varepsilon-\varepsilon_{c}}{\Lambda-1}\exp\left[
\frac{\chi-\nu\tau}{\left\vert \xi_{\Lambda,-}\right\vert }\right]
\qquad\qquad & \text{{for}}\,\,\chi-\nu\tau<0\quad\text{({rear side)}}\\
& \\
\displaystyle\varepsilon-\frac{\varepsilon_{c}-\varepsilon}{\Lambda-1}%
\exp\left[  -\frac{\chi-\nu\tau}{\xi_{1,+}}\right]  & \text{{for}}%
\,\,\,0\leq\chi-\nu\tau\quad\text{{(front side)}}.
\end{cases}
\end{equation}
Here, two healing lengths have been introduced (see Fig.~4e in the main text and Fig.~\ref{fig:sup:HealingLength}c below):
\begin{equation}
\left\vert \xi_{\Lambda,-}\right\vert =\sqrt{\frac{\Lambda\tilde{\varepsilon}-1}{\tilde{\varepsilon}(1-\tilde{\varepsilon})}},
\qquad
\xi_{1,+}=\sqrt{\frac{1-\tilde{\varepsilon}}{\tilde{\varepsilon}(\Lambda\tilde{\varepsilon}-1)}}.
\end{equation}
Equation~(\ref{eq:sup:nuepsilon_for_KV}) or (\ref{eq:sup:epsilonnu_for_KV}) gives the relation 
between the velocity $V$ and the initially applied energy density $w=\frac{1}{2}E_{0}\varepsilon^{2}$, by virtue of the
relation
\begin{equation}
\frac{w}{\Lambda w_{0}}
=\tilde{\varepsilon}^{2}\equiv\frac{\varepsilon^{2}}{\varepsilon_{c}^{2}}
\text{ \quad with \quad}w_{0}=\frac{E_{0}\varepsilon_{c}^{2}}{2\Lambda}. 
\label{eq:sup:w-epsilon}
\end{equation}

\newpage
\section{Minimal model consisting of Zener elements 1: the basic equations
and exact solution}

\begin{figure}[h]
\centering
\includegraphics[keepaspectratio, scale=1.0]{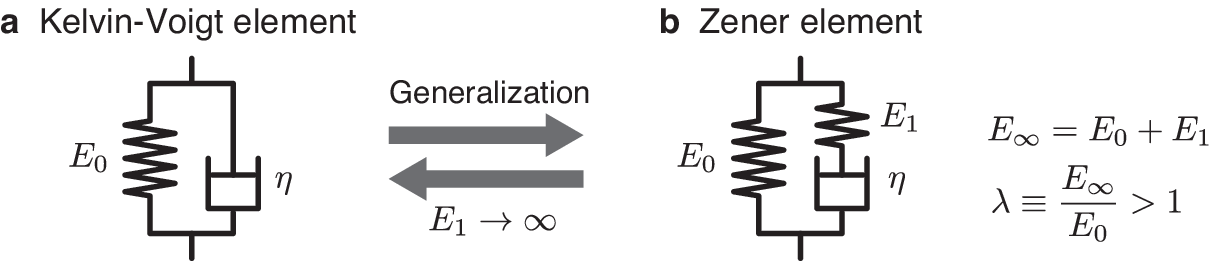}
\caption{\textbf{(a) Kelvin-Voigt and (b) Zener elements. }
We obtain the former as the limit $E_{1} \to \infty$ from the latter.}
\label{fig:sup:VoigtZener}
\end{figure}

\subsection{Generalization of viscoelastic interaction: equations of motion in the $y$-direction}

In this section, we generalize the present model by changing the interaction
from the one based on Kelvin-Voigt elements to the one on Zener elements
(see Fig.~\ref{fig:sup:VoigtZener}).
A Zener element is a parallel connection of the spring $E_{0}$ (strain
$\mathcal{E}(t)$) and a Maxwell element, which is a serial connection of the
spring $E_{1}$ (strain $\mathcal{E}_{1}(t)$) and the dashpot $\eta$ (strain
$\mathcal{E}_{2}(t)$). 
The strains of the parallel components are identical:
\begin{equation}
\mathcal{E}(t)=\mathcal{E}_{1}(t)+\mathcal{E}_{2}(t).
\label{eq:sup:EE1E2}%
\end{equation}
For the serial components in the Maxwell element, the stresses on the spring
$E_{1}$ and dashpot $\eta$ are identical:
\begin{equation}
\eta\frac{\partial}{\partial t}\mathcal{E}_{2}(t)=E_{1}\mathcal{E}_{1}(t).
\label{eq:sup:E1E2}
\end{equation}
Note here that a Zener element reduces to a Kelvin-Voigt element 
 in the limit $E_{1}\rightarrow\infty$, in which $\mathcal{E}_{1}(t)=0$.
The generalization of the equations of motion in the $y$-direction is achieved
by the following replacements:
\begin{equation}%
\begin{cases}
\displaystyle\frac{F_{\mathrm{up}}}{ah}+\frac{F_{\mathrm{up}}^{(\eta)}}{ah}
=E_{0}\mathcal{E}+\eta\frac{\partial}{\partial t}\mathcal{E} & \displaystyle\rightarrow\quad
E_{0}\mathcal{E}+\eta\frac{\partial}{\partial t}\mathcal{E}_{2}\\
& \\
\displaystyle\frac{F_{\mathrm{up}}}{ah}+\frac{F_{\mathrm{up}}^{(\eta)}}{ah}-\frac{F_{\mathrm{down}}}{ah}-\frac{F_{\mathrm{down}}^{(\eta)}}{ah}
=\frac{L}{l}E_{0}\left(\mathcal{E}-\varepsilon\right)+\frac{L}{l}\eta\frac{\partial}{\partial t}\mathcal{E}\quad & \displaystyle\rightarrow
\quad\frac{L}{l}E_{0}\left(\mathcal{E}-\varepsilon\right)+\frac{L}{l}\eta\frac{\partial}{\partial t}\mathcal{E}_{2}.
\end{cases}
\end{equation}
With this replacement, Eq.~(\ref{eq:sup:EoSperAH}) in the continuum limit in the $x$-direction, $a\to 0$,
is given as
\begin{equation}%
\begin{cases}
\displaystyle0=E_{0}\mathcal{E}+\eta\frac{\partial}{\partial t}\mathcal{E}_{2}-\frac{1}{2}\mu l(L-l)\frac{\partial^{2}}{\partial x^{2}}\mathcal{E}
 & \text{rear side}\\
& \\
\displaystyle
0=\frac{L}{l}E_{0}\left(  \mathcal{E}-\varepsilon\right)
+\frac{L}{l}\eta\frac{\partial}{\partial t}\mathcal{E}_{2}-\frac{1}{2}\mu l(L-l)\frac{\partial^{2}}{\partial x^{2}}\mathcal{E}\qquad 
& \text{front side}
\end{cases}
\end{equation}
As in the case of the Kelvin-Voigt interaction, 
we use the dimensionless parameters, 
$\tau=t/t_{0}$, $\chi=x/x_{0}$, and $\Lambda \equiv L/l$,
where $t_{0} \equiv \eta/E_{0}$, and $x_{0}\equiv l\sqrt{(L-l)\mu /(2LE_{0})}$.
Then, we rewrite the above equations of motion as
\begin{equation}
\begin{cases}
\displaystyle0=\mathcal{E}+\dot{\mathcal{E}}_{2}-\Lambda\mathcal{E}^{\prime\prime}
 & \text{rear side}\\
\displaystyle0=\left(  \mathcal{E}-\varepsilon\right)  +\dot{\mathcal{E}}_{2}-\mathcal{E}^{\prime\prime}\qquad
 & \text{front side,}
\end{cases}
\label{eq:sup:EoM-Zener-EE2}%
\end{equation}
where $\dot{\mathcal{E}}_{2}\equiv\frac{\partial}{\partial\tau}\mathcal{E}_{2}(\tau,\chi)$ 
and $\mathcal{E}^{\prime\prime}\equiv\frac{\partial^{2}}{\partial\chi^{2}}\mathcal{E}(\tau,\chi)$. 

We eliminate $\mathcal{E}$ in Eq.~(\ref{eq:sup:EoM-Zener-EE2}) 
by using the following relation obtained by substituting Eq.~(\ref{eq:sup:E1E2}) into
Eq.~(\ref{eq:sup:EE1E2}),
\begin{equation}
\mathcal{E}=\mathcal{E}_{2}+\frac{E_{0}}{E_{1}}\dot{\mathcal{E}}_{2},
\label{eq:sup:EE2dotE2}%
\end{equation}
to have the equations of motion for a single field
$\mathcal{E}_{2}$ in the following form:
\begin{equation}%
\begin{cases}
\displaystyle0=\mathcal{E}_{2}+(1+s)\dot{\mathcal{E}}_{2}-\Lambda
\mathcal{E}_{2}^{\prime\prime}-s\Lambda\dot{\mathcal{E}}_{2}^{\prime\prime} &
\text{{rear side}}\\
\displaystyle0=\left(  \mathcal{E}_{2}-\varepsilon\right)
  +(1+s)\dot{\mathcal{E}}_{2}-\mathcal{E}_{2}^{\prime\prime}-s\dot{\mathcal{E}}_{2}^{\prime\prime}
  \qquad & \text{front side}.
\end{cases}
\label{eq:sup:EoM-Zener-E2}%
\end{equation}
Here, we have introduced the parameter
\begin{equation}
s\equiv\frac{E_{0}}{E_{1}}.
\label{eq7}%
\end{equation}
Equation~(\ref{eq:sup:EoM-Zener-E2}) with $\mathcal{E}_{2}$ replaced by $\mathcal{E}$ also holds.
This equation for $\mathcal{E}$ can be obtained either (i) by multiplying the operator $1+s\frac{\partial}{\partial\tau}$ to Eq.~(\ref{eq:sup:EoM-Zener-E2})
and use $(1+s\frac{\partial}{\partial\tau})\mathcal{E}_{2}=\mathcal{E}$
or (ii) by directly eliminating $\mathcal{E}_{2}$ in Eq.~(\ref{eq:sup:EoM-Zener-EE2}).

In the main text, we have used $\lambda$ defined as
\begin{equation}
\lambda\equiv \frac{E_{\infty}}{E_{0}}=1+\frac{1}{s},
\end{equation}
instead of $s$.
Although the physical meaning of $\lambda$ is clearer than $s$,
we mainly use $s$ in this section for mathematical simplicity.
The parameters $s$ and $\lambda$ vary in the ranges of 
$0< s<\infty$ and $1<\lambda<\infty$, respectively.
Note that the Kelvin-Voigt interaction (finite $E_{0}$ and $E_{1}=\infty$) corresponds
to the limit $s \to 0$ or equivalently $\lambda \to \infty$.

\subsection{Derivation of the exact solution of crack propagation with a constant velocity}

In this subsection, we solve Eq.~(\ref{eq:sup:EoM-Zener-E2}) in the case of crack propagation with a constant velocity ($0<V<\infty$).
Note that, in the case of $V=0$, a Zener element reduces to a Kelvin-Voigt element,
and we have already solved the model consisting of Kelvin-Voigt elements in Sec.~\ref{sec:sup:KV}.

We derive the equations of motion with relevant boundary conditions to be
solved in the present case, following the manner employed in the Kelvin-Voigt case in Sec.~\ref{sec:sup:KV}. 
We substitute the form $\mathcal{E}_{2}(\tau,\chi
)=f(\chi-\nu\tau)$ into Eq.~(\ref{eq:sup:EoM-Zener-E2}) and derive linear ordinary differential equations for $f(\chi)$ for $\chi<0$ and $0\leq\chi$, where we set the position of the crack tip to $\chi =0$ with $\tau =0$.
Here, $V$ and
$\nu=V/V_{0}$ are the dimensional and dimensionless velocities, respectively,
with $V_{0}\equiv x_{0}/t_{0}\simeq lE_{0}/\eta$ as before 
(see Eq.~(\ref{eq:sup:def:V0})). 
We give the result as
\begin{equation}
\begin{cases}
\displaystyle0=f(\chi)-(1+s)\nu f^{\prime}(\chi)-\Lambda f^{\prime\prime}%
(\chi)+s\nu\Lambda f^{\prime\prime\prime}(\chi)\qquad & \text{{for}}%
\,\,\chi<0\quad\text{{(rear side)}}\\
\displaystyle0=f(\chi)-\varepsilon-(1+s)\nu f^{\prime}(\chi)-f^{\prime\prime
}(\chi)+s\nu f^{\prime\prime\prime}(\chi) & \text{{for}}\,\,\,0\leq\chi
\quad\text{{(front side)}}.
\end{cases}
\label{eq:sup:eom-Zener-steady}
\end{equation}

To determine the boundary conditions, 
we substitute $\mathcal{E}_{2}(\tau,\chi)=f(\chi-\nu\tau)$ into Eq.~(\ref{eq:sup:EE2dotE2}) 
to have $\mathcal{E}=f-s\nu f^{\prime}$.
For $\chi-\nu\tau\rightarrow\pm\infty$, 
we have $\mathcal{E}=f$, since $f^{\prime}=0$. 
For $\chi-\nu\tau=0$, 
we assume that the strain distributions 
$\mathcal{E}$ and $\mathcal{E}_{2}$
are continuous and differentiable at the crack tip.
The values of the functions and their derivatives match at the crack tip: 
$\mathcal{E}(-0)=\mathcal{E}(+0)$,
$\mathcal{E}^{\prime}(-0)=\mathcal{E}^{\prime}(+0)$, and
$\mathcal{E}_{2}(-0)=\mathcal{E}_{2}(+0)$
(we can derive 
$\mathcal{E}_{2}^{\prime}(-0)=\mathcal{E}_{2}^{\prime}(+0)$ from the others). 
In addition, as in the case of the Kelvin-Voigt interaction, we
assume that the short spring is absent when $\mathcal{E}^{(d)}\geq
\varepsilon_{c}$, which yields $\mathcal{E}(-0)=\mathcal{E}(+0)=\frac
{\Lambda\varepsilon-\varepsilon_{c}}{\Lambda-1}$. 
In summary, we give the appropriate boundary conditions as
\begin{equation}
\begin{cases}
\displaystyle f(-0)-s\nu f^{\prime}(-0)=f(+0)-s\nu f^{\prime}(+0)=\frac
{\Lambda\varepsilon-\varepsilon_{c}}{\Lambda-1}; & \\
\displaystyle f^{\prime}(-0)=f^{\prime}(+0); & f^{\prime\prime}(-0)=f^{\prime
\prime}(+0);\\
\displaystyle f(-\infty)=0; & f(+\infty)=\varepsilon.
\end{cases}
\label{eq:sup:bc-Zener-steady}
\end{equation}

We can then establish the exact analytical relation between the global strain $\varepsilon$ and the normalized velocity of
crack propagation $\nu\equiv V/V_{0}$ as in the following theorem:

\newpage

\textbf{Theorem~1 (Relation between global strain and crack-propagation velocity).} 
\textit{
If Eqs.~(\ref{eq:sup:eom-Zener-steady}) and
(\ref{eq:sup:bc-Zener-steady}) hold, then
\begin{equation}
\tilde{\varepsilon}
\equiv\frac{\varepsilon}{\varepsilon_{c}}
= \frac{\nu\left(1+s+\frac{s\Lambda}{\xi_{1}\xi_{\Lambda}}\right)  +\xi_{1}+\xi_{\Lambda}}
          {\nu\left(1+s\Lambda+\frac{s\Lambda}{\xi_{1}\xi_{\Lambda}}\right)
                +\Lambda\xi_{1}+\xi_{\Lambda}}
= \frac{\frac{\nu}{\lambda-1}\left(\frac{\Lambda}{\xi_{1}\xi_{\Lambda}}+\lambda\right)
                +\xi_{1}+\xi_{\Lambda}}
         {\frac{\nu}{\lambda-1}\left(  \Lambda+\frac{\Lambda}{\xi_{1}\xi_{\Lambda}}+\lambda-1\right)
                +\Lambda\xi_{1}+\xi_{\Lambda}},
\label{eq:sup:thm:epsilonnu}
\end{equation}
where
}
\begin{equation}
\begin{split}
\xi_{\Lambda} &  =\frac{1}{6}\Biggl[2^{2/3}\sqrt[3]{3\sqrt{3}\sqrt
{-\Lambda\left(  4\Lambda^{2}+\Lambda(4(5-2s)s+1)\nu^{2}+4s(s+1)^{3}\nu
^{4}\right)  }+9\Lambda(2s-1)\nu-2(s+1)^{3}\nu^{3}}\\
&  +\frac{2\sqrt[3]{2}\left(  3\Lambda+(s+1)^{2}\nu^{2}\right)  }%
{\sqrt[3]{3\sqrt{3}\sqrt{-\Lambda\left(  4\Lambda^{2}+\Lambda(4(5-2s)s+1)\nu
^{2}+4s(s+1)^{3}\nu^{4}\right)  }+9\Lambda(2s-1)\nu-2(s+1)^{3}\nu^{3}}%
}-2(s+1)\nu\Biggr].
\end{split}
\label{eq:sup:thm:xiLambda}
\end{equation}\\

We note that $\xi_{\Lambda}$ is the positive solution of the cubic equation
\begin{equation}
g_{\Lambda}(\xi)\equiv\xi^{3}+(1+s)\nu\xi^{2}-\Lambda\xi-s\nu\Lambda
=0.
\label{eq:sup:def:gLambda}
\end{equation}
In the limit $s\rightarrow0$,
$g_{\Lambda}(\xi)=0$ reduces
to the quadratic equation~(\ref{eq:sup:def:gLambda-KVoigt})
appearing in the case of the Kelvin-Voigt interaction. 
In this limit, 
Eqs.~(\ref{eq:sup:thm:epsilonnu}) and (\ref{eq:sup:thm:xiLambda})
reduce to Eq.~(\ref{eq:sup:epsilonnu_for_KV}) and
$\xi_\Lambda = \left( \sqrt{\nu^2+4\Lambda} -\nu \right)/2$,
respectively.
The uniqueness of $\xi_{\Lambda}$ is ensured by the following lemma.\\

\begin{lemma}
Let $s$, $\nu$, and $\Lambda$ are positive real numbers. Then,
$g_{\Lambda}(\xi) =0$ has one positive and two negative real solutions for
$-\infty< \xi< \infty$. 
\end{lemma}


We prove Theorem~1 with the aid of Lemma~1 
(we give the proof of Lemma~1 at the last of this subsection). \\

\noindent
\textit{Proof of Theorem~1.}

First, we solve the differential equation for $\chi<0$ (the first equation of
Eqs.~(\ref{eq:sup:eom-Zener-steady})). Substituting the form $f(\chi)=Ce^{-\chi
/\xi}$ into the equation, we obtain the characteristic
equation~(\ref{eq:sup:def:gLambda}). 
According to Lemma~1, we can set the
three solutions, $\xi_{\Lambda,1}$, $\xi_{\Lambda,2}$, and $\xi_{\Lambda}$, of
the cubic equation $g_{\Lambda}(\xi)=0$ to satisfy the relation
$\xi_{\Lambda,1}<\xi_{\Lambda,2}<0<\xi_{\Lambda}$.
To satisfy the boundary conditions at $\xi\rightarrow-\infty$, 
the solution of the differential equation for $\chi<0$ has the form
$f(\chi)=\sum_{i=1}^{2}C_{i}e^{-\chi/\xi_{\Lambda,i}}$, 
where $C_{1}$ and $C_{2}$ will be determined later by using the boundary conditions at $\chi=0$.

Second, we solve the differential equation for $\chi\geq0$. Substituting the form
$f(\chi)-\varepsilon=Ce^{-\chi/\xi}$ into the equation, we obtain the
characteristic equation,
\begin{equation}
g_{1}(\xi)=\xi^{3}+(1+s)\nu\xi^{2}-\xi-s\nu=0.
\end{equation}
Note that $g_{1}(\xi)$ is identical to $g_{\Lambda}(\xi)$ with $\Lambda=1$.
According to Lemma~1, $g_{1}(\xi)=0$ has one positive and two negative real
solutions, and we denote the positive solution as $\xi_{1}$. 
To satisfy the boundary conditions at $\xi\rightarrow\infty$, 
the solution of the differential equation for $\chi>0$ has the form
$f(\chi)=\varepsilon-C_{0}e^{-\chi/\xi_{1}}$, 
where we can determine $C_{0}$ by using the boundary condition at $\chi=0$:
\begin{equation}
\frac{\Lambda\varepsilon-\varepsilon_{c}}{\Lambda-1}=f(+0)-s\nu f^{\prime
}(+0)=\varepsilon-C_{0}\left(  1+\frac{s\nu}{\xi_{1}}\right)  .
\end{equation}

Finally, we determine the relation between $\tilde{\varepsilon}$ and $\nu$, by
eliminating $C_{0}$, $C_{1}$, and $C_{2}$
from the boundary conditions,
$f(-0)=f(+0)$, $f^{\prime}(-0)=f^{\prime}(+0)$, and $f^{\prime\prime}(-0)=f^{\prime\prime}(+0)$,
which can be recast into the following forms:
\begin{equation}%
\begin{cases}
\varepsilon-C_{0}=C_{1}+C_{2}\\
\xi_{1}^{-1}C_{0}=-\xi_{\Lambda,1}^{-1}C_{1}-\xi_{\Lambda,2}^{-1}C_{2}\\
-\xi_{1}^{-2}C_{0}=\xi_{\Lambda,1}^{-2}C_{1}+\xi_{\Lambda,2}^{-2}C_{2}.
\end{cases}
\label{eq:sup:three_previous}%
\end{equation}
Introducing the parameters 
$D_{i}\equiv C_{i}/C_{0}$ 
and 
$\gamma_{i}\equiv\left\vert \xi_{1}/\xi_{\Lambda,i}\right\vert =-\xi_{1}/\xi_{\Lambda,i}$, 
Eq.~(\ref{eq:sup:three_previous}) is written as
\begin{equation}%
\begin{cases}
\displaystyle D_{1}+D_{2}=\frac{\varepsilon}{C_{0}}-1=\left(  1+\frac{s\nu
}{\xi_{1}}\right)  \frac{\Lambda-1}{\tilde{\varepsilon}^{-1}-1}-1\\
\gamma_{1}D_{1}+\gamma_{2}D_{2}=1\\
\gamma_{1}^{2}D_{1}+\gamma_{2}^{2}D_{2}=-1.
\end{cases}
\label{eq:sup:three}
\end{equation}
From the second and third expressions of Eq.~(\ref{eq:sup:three}), we obtain
$D_{1}=\frac{\gamma_{2}+1}{\gamma_{1}(\gamma_{2}-\gamma_{1})}$ and
$D_{2}=\frac{-1-\gamma_{1}}{\gamma_{2}(\gamma_{2}-\gamma_{1})}$, 
which leads
\begin{equation}
D_{1}+D_{2}=\frac{1}{\gamma_{1}}+\frac{1}{\gamma_{2}}+\frac{1}{\gamma
_{1}\gamma_{2}}.\label{eq:sup:D1D2}%
\end{equation}
Here, we note that  $\gamma_{1}<\gamma_{2}$.
According to Vieta's Formulae (which is for a polynomial equation and relates
the coefficients of the polynomial with sums and products of the roots), the
characteristic equation $g_{\Lambda}(\xi)=0$ gives $\xi_{\Lambda,1}%
+\xi_{\Lambda,2}+\xi_{\Lambda}=-(1+s)\nu$ and $\xi_{\Lambda,1}\xi_{\Lambda
,2}\xi_{\Lambda}=s\nu\Lambda$.
These two relations are rewritten as
\begin{equation}%
\begin{cases}
\displaystyle\frac{1}{\gamma_{1}}+\frac{1}{\gamma_{2}}=-\frac{\xi_{\Lambda,1}%
}{\xi_{1}}-\frac{\xi_{\Lambda,2}}{\xi_{1}}=\frac{1}{\xi_{1}}\left[
(1+s)\nu+\xi_{\Lambda}\right]  \\
\\
\displaystyle\frac{1}{\gamma_{1}\gamma_{2}}=\frac{\xi_{\Lambda,1}\xi
_{\Lambda,2}}{\xi_{1}^{2}}=\frac{s\nu\Lambda}{\xi_{1}^{2}\xi_{\Lambda}}.
\end{cases}
\label{eq:sup:VietaFormulas}%
\end{equation}
Substituting Eqs.~(\ref{eq:sup:VietaFormulas}) into Eq.~(\ref{eq:sup:D1D2}),
we have
\begin{equation}
D_{1}+D_{2}=\frac{1}{\xi_{1}}\left[  (1+s)\nu+\xi_{\Lambda}\right]
+\frac{s\nu\Lambda}{\xi_{1}^{2}\xi_{\Lambda}}.
\end{equation}
Combining this with the first relation in Eq.~(\ref{eq:sup:three}), we have
Eq.~(\ref{eq:sup:thm:epsilonnu}). Thus, Theorem~1 is proved. (Q.E.D.) \\

\begin{table}[t]
\caption{The behaviors of $g_{\Lambda}(\xi)$ as a function of $\xi$ (left) and
$g_{\Lambda}(\alpha_{-})$ as a function of $\Lambda$ (right).}%
\label{tab1}%
\begin{tabular}
[c]{ll}%
\begin{tabular}
[c]{c|ccccccc}\hline\hline
$\xi$ & \qquad\qquad\qquad & $\alpha_{-}$ & \qquad\qquad\qquad & $0$ &
\qquad\qquad\qquad & $\alpha_{+}$ & \qquad\qquad\qquad\\\hline
$\frac{\partial}{\partial\xi} g_{\Lambda}(\xi)$ & $+$ & $0$ & $-$ & $-$ & $-$
& $0$ & $+$\\
$g_{\Lambda}(\xi)$ & $\nearrow$ & $g_{\Lambda}(\alpha_{-})$ & $\searrow$ &
$-s\nu\Lambda$ & $\searrow$ & $g_{\Lambda}(\alpha_{+})$ & $\nearrow
$\\\hline\hline
\end{tabular}
\qquad & \qquad%
\begin{tabular}
[c]{c|ccccccc}\hline\hline
$\Lambda$ & \qquad\qquad\qquad & $s(s-2)\nu^{2}$ & \qquad\qquad\qquad &  &  &
& \\\hline
$\frac{\partial}{\partial\Lambda} g_{\Lambda}(\alpha_{-}) $ & $-$ & $0$ & $+$
&  &  &  & \\
$g_{\Lambda}(\alpha_{-}) $ & $\searrow$ & $s^{2}\nu^{3}$ & $\nearrow$ &  &  &
& \\\hline\hline
\end{tabular}
\end{tabular}
\end{table}

\noindent
\textit{Proof of Lemma~1.} 

We express the equation $\frac{\partial}{\partial\xi}g_{\Lambda}(\xi)=0$ as $3\xi^{2}+2(1+s)\nu\xi-\Lambda =0$ and denote its two solutions as 
$\alpha_{\pm}=\left[  -(1+s)\nu\pm\sqrt{\nu^{2}(1+s)^{2}+3\Lambda}\right]  /3$, 
where $\alpha_{-}<0<\alpha_{+}$. 
The behavior of
$g_{\Lambda}(\xi)$ as a function of $\xi$ is summarized in Tab.~\ref{tab1};
for $\alpha_{-}\leq\xi\leq\alpha_{+}$, the function $g_{\Lambda}(\xi)$ is monotonically decreasing. 
Then, $g_{\Lambda}(\alpha_{+})<g_{\Lambda}(0)=-s\nu\Lambda<0$. 
If $g_{\Lambda}(\alpha_{-})>0$, $g_{\Lambda}(\xi)=0$ has
one positive and two negative real solutions. Thus, in the following, we show
$g_{\Lambda}(\alpha_{-})>0$.

To show $g_{\Lambda}(\alpha_{-})>0$, we calculate $g_{\Lambda}(\alpha_{-})$
and its derivative as follows:
\begin{align}
&  g_{\Lambda}(\alpha_{-})=\frac{2}{27}\left[  3\Lambda+(s+1)^{2}\nu
^{2}\right]  ^{3/2}+\frac{\Lambda\nu}{3}\left(  1-2s\right)  +\frac{2}%
{27}(s+1)^{3}\nu^{3},\\
&  \frac{\partial}{\partial\Lambda}g_{\Lambda}(\alpha_{-})=\frac{1}{3}\left[
3\Lambda+(s+1)^{2}\nu^{2}\right]  ^{1/2}+\frac{\nu}{3}\left(  1-2s\right)  ,\\
&  \frac{\partial^{2}}{\partial\Lambda^{2}}g_{\Lambda}(\alpha_{-})=\frac{1}%
{2}\left[  3\Lambda+(s+1)^{2}\nu^{2}\right]  ^{-1/2}.
\end{align}
Since $s$, $\nu$, and $\Lambda$ are positive real numbers, $\frac{\partial
^{2}}{\partial\Lambda^{2}}g_{\Lambda}(\alpha_{-})>0$, which means that
$\frac{\partial}{\partial\Lambda}g_{\Lambda}(\alpha_{-})$ is a monotonically
increasing function. 
Now, $\frac{\partial}{\partial\Lambda}g_{\Lambda}%
(\alpha_{-})=0$ at $\Lambda=s(s-2)\nu^{2}$, 
at which $g_{\Lambda}(\alpha_{-})$ takes the minimum value as a function of $\Lambda$.
Since $g_{\Lambda}(\alpha_{-})|_{\Lambda=s(s-2)\nu^{2}}=s^{2}\nu^{3}$ is positive, $g_{\Lambda}(\alpha_{-})>0$ for any $\Lambda$. Thus, Lemma~1 is proved. (Q.E.D.) \\

\newpage
\section{Minimal model consisting of Zener elements 2: low- and high-velocity regimes and velocity jump}

In this section, 
we use Eq.~(\ref{eq:sup:thm:epsilonnu}) 
to investigate the dependences of the initially applied energy density $w$ on the velocity $\nu\equiv V/V_{0}$ 
in low- and high-velocity regimes, and derive the existence condition of the velocity jump. 
Note that Eq.~(\ref{eq:sup:thm:epsilonnu}) gives the relation between the initially applied global strain $\varepsilon$ and $\nu$
and that $\varepsilon$ is related to $w$ simply as $w=\tilde{\varepsilon}^2 \Lambda w_{0}$ 
(see Eq.~(\ref{eq:sup:w-epsilon})).
In the following, the present model will be analyzed for arbitrary positive real number $s\equiv E_{0}/E_{1} >0$, 
which is equivalent to
$\lambda\equiv E_{\infty}/E_{0}=1+1/s >1$. 
Note that the relations derived below are further simplified for elastomers, 
because of the relation $1 \ll \lambda \ll \Lambda$
(typically $\lambda \simeq 10^{2}$--$10^{3}$ and $\Lambda \simeq 10^{6}$--$10^{9}$).

\subsection{Low- and high-velocity regimes}

To obtain the asymptotic forms of the initially applied energy $w$ in the low- and high-velocity regimes, 
we evaluate that of the characteristic equation~(\ref{eq:sup:def:gLambda}).
We rewrite this equation as
\begin{equation}
\frac{g_{\Lambda}(\xi)}{\Lambda^{3/2}}
=\Xi^{3}
+\frac{\lambda\upsilon}{\lambda -1}\Xi^{2}
-\Xi
-\frac{\upsilon}{\lambda -1}=0,
\label{eq:sup:def:gLambdaOverLambda}
\end{equation}
where 
$\Xi\equiv\xi/\sqrt{\Lambda}$ and $\upsilon\equiv\nu/\sqrt{\Lambda}$. 
We denote the positive real
solution of Eq.~(\ref{eq:sup:def:gLambdaOverLambda}) as $\Xi_{+}\equiv
\xi_{\Lambda}/\sqrt{\Lambda}$, where $\xi_{\Lambda}$ is the positive real
solution of Eq.~(\ref{eq:sup:def:gLambda}). We note that $\Xi_{+}$ is
independent of $\Lambda$ and depends only on $\upsilon$ $(0<\upsilon<\infty)$
and $s$, as seen from Eq.~(\ref{eq:sup:def:gLambdaOverLambda}). 
Then, we give the relation between $\xi_{\Lambda}$ and $\xi_{1}$ as 
$\Xi_{+}(\upsilon)=\xi_{\Lambda}(\nu/\sqrt{\Lambda})/\sqrt{\Lambda}=\xi_{1}(\nu)$.
By evaluating the asymptomatic forms of $\Xi_{+}$ 
in the low- and high-velocity regimes, 
we have the following lemma:\\


\begin{lemma}
If $0<\upsilon\equiv\nu/\sqrt{\Lambda}<\infty$ and $1<\lambda<\infty$, then
the positive real solution $\xi_{\Lambda}$ of the
characteristic equation~(\ref{eq:sup:def:gLambda}), $g_{\Lambda}(\xi)=0$, has
the following asymptotic forms:
\begin{equation}%
\frac{1}{\sqrt{\Lambda}} \, \xi_{\Lambda}(\nu)
\equiv \Xi_{+} (\upsilon)
=
\begin{cases}
\displaystyle1-\frac{\upsilon}{2}+O\left(  \upsilon^{2}\right)   & (\upsilon\to 0)\\
\displaystyle
\frac{1}{\sqrt{\lambda}}
+\frac{(\lambda-1)^{2}}{2\upsilon\lambda^{2}}
+O\left({\frac{1}{\upsilon^{2}}}\right)
 & (\upsilon\to\infty).
\end{cases}
\end{equation}
\end{lemma}

Note here that (i) if $\nu \equiv V/V_0 \ll\sqrt{\Lambda}$, then $\xi_{\Lambda}(\nu)\simeq\sqrt{\Lambda}$; 
and (ii) if
$\nu\gg\sqrt{\Lambda}(\lambda-1)^{2}/\lambda^{3/2}$,
then $\xi_{\Lambda}(\nu)\simeq\sqrt{\Lambda/\lambda}$.\\

\noindent
\textit{Proof of Lemma~2.} 

We evaluate the asymptomatic form of $\Xi_{+}$ in the vicinity of $\upsilon=0$
(i.e., in the low-velocity regime). 
Because of the asymptotic behavior,
$g_{\Lambda}(\xi)/\Lambda^{3/2}\xrightarrow[\upsilon \to 0]{}\Xi^{3}-\Xi$,
We can express Eq.~(\ref{eq:sup:def:gLambdaOverLambda}) as $\Xi^{3}-\Xi=0$, 
which gives the positive real solution, $\Xi=1$. 
To evaluate the next order, we
introduce the perturbation parameter $\epsilon_{0}$. Substituting
$\Xi=1+\epsilon_{0}$ into Eq.~(\ref{eq:sup:def:gLambdaOverLambda}), we have
\begin{equation}
2\epsilon_{0}+\upsilon+O(\epsilon_{0}^{2}, \epsilon_{0}\upsilon)=0,
\end{equation}
whose solution is $\epsilon_{0}=-\upsilon/2+O(\upsilon^{2})$. 
Therefore, we have the expression, $\Xi_{+}=1-\upsilon/2+O(\upsilon^{2})$.

Similarly, we evaluate the asymptomatic form of $\Xi_{+}$ in the vicinity of
$1/\upsilon=0$ (i.e., in the high-velocity regime, $\upsilon\gg1$). 
Because of the asymptotic behavior,
$g_{\Lambda}(\xi)\cdot(\lambda-1)/(\upsilon\Lambda^{3/2})
\xrightarrow[\upsilon\to \infty]{}
\lambda\Xi^{2}-1$,
Eq.~(\ref{eq:sup:def:gLambdaOverLambda}) can be expressed as $\lambda\Xi^{2}-1=0$,
which gives the positive real solution, $\Xi=1/\sqrt{\lambda}$.
Introducing the perturbation parameter $\epsilon_{\infty}$, and substituting
$\Xi=1/\sqrt{\lambda}+\epsilon_{\infty}$ into
Eq.~(\ref{eq:sup:def:gLambdaOverLambda}), we have
\begin{equation}
\frac{2\sqrt{\lambda}}{\lambda-1}\upsilon\epsilon_{\infty}
-\frac{\lambda-1}{\lambda^{3/2}}
- \frac{\lambda -3}{\lambda}\epsilon_{\infty}
+\frac{\lambda}{\lambda-1}\upsilon\epsilon_{\infty}^{2}
+\frac{3}{\sqrt{\lambda}}\epsilon_{\infty}^{2}
+\epsilon_{\infty}^{3}
=0.
\label{eq:sup:proof:lemma2:high-velocity}
\end{equation}
Since $1<\lambda<\infty$, $\left|\epsilon_{\infty}\right| \ll1$, and $\upsilon^{-1} \ll1$, 
the first and second terms on the left-hand side of Eq.~(\ref{eq:sup:proof:lemma2:high-velocity})
are the leading-order terms.
Thus, the solution of Eq.~(\ref{eq:sup:proof:lemma2:high-velocity}) is 
$\epsilon_{\infty}
=\frac{1}{2\upsilon}\left(  \frac{\lambda-1}{\lambda}\right)^{2}
+O\left(  \upsilon^{-2}\right)$.
Therefore, we have the expression, 
$\Xi_{+}
=\frac{1}{\sqrt{\lambda}}
+\frac{1}{2\upsilon}\left(\frac{\lambda-1}{\lambda}\right)^{2}
+O\left(\upsilon^{-2}\right)$.
(Q.E.D)\\

From Theorem~1, with the aid of Lemma~2, we can evaluate
$\tilde{\varepsilon}(\nu)$ in the vicinity of $\nu=0$ and that of $\nu=\infty$. 
In the vicinity of $\nu=0$, 
$\xi_{\Lambda}(\nu)=\sqrt{\Lambda}-\frac{\nu}{2}+O(\nu^{2})$ and $\xi_{1}(\nu)=1-\frac{\nu}{2}+O(\nu^{2})$. 
Then, together with Eq.~(\ref{eq:sup:thm:epsilonnu}), 
we can evaluate $\tilde{\varepsilon}(\nu)$.
Similarly, in the vicinity of $\nu=\infty$, 
$\xi_{\Lambda}(\nu)=
\sqrt{\frac{\Lambda}{\lambda}}
+\frac{\Lambda(\lambda-1)^{2}}{2\nu\lambda^{2}}
+O\left(  \frac{1}{\nu^{2}}\right)  $ and 
$\xi_{1}(\nu)
=\frac{1}{\sqrt{\lambda}}
+\frac{(\lambda-1)^{2}}{2\nu\lambda^{2}}
+O\left(  \frac{1}{\nu^{2}}\right)$. 
Then, together with Eq.~(\ref{eq:sup:thm:epsilonnu}) we can evaluate $\tilde{\varepsilon}(\nu)$.
We summarize the evaluation in the following theorem.\\

\textbf{Theorem~2 (Asymptotic forms in low- and high-velocity regimes).} 
\textit{
If $\lambda >1$ and $\Lambda >1$, then
}
\begin{equation}
\tilde{\varepsilon} (\nu)
 = \frac{1}{\sqrt{\Lambda}}
  + \frac{\sqrt{\Lambda}-1}{2\Lambda} \nu+ O\left(  \nu^{2} \right)  \qquad(\nu\to 0),
\label{eq:sup:thm:AsymptoticLowV}
\end{equation}
\textit{and}
\begin{equation}%
\begin{split}
\tilde{\varepsilon} (\nu)
&  = \frac{\lambda}{\sqrt{\Lambda} + \lambda-1}
 - \frac
    {   \left(\lambda-1\right)^{2}\left(  \sqrt{\Lambda}-1\right) \left(  \sqrt{\Lambda}+2\right)}
    { 2\nu \sqrt{\lambda} \left(  \sqrt{\Lambda} + \lambda-1\right)^{2} }
+ O\left(  \frac{1}{\nu^{2}}\right)
\end{split}
\qquad(\nu\to\infty).
\label{eq:sup:thm:AsymptoticHighV}
\end{equation}\\

Equation~(\ref{eq:sup:thm:AsymptoticLowV}) enables us to evaluate 
the relation between the initially applied energy density $w$ and the crack-propagation velocity $V$
in the low-velocity regime
(by virtue of the relation $w=\tilde{\varepsilon}^2 \Lambda w_{0}$, see Eq.~(\ref{eq:sup:w-epsilon})).
From Eq.~(\ref{eq:sup:thm:AsymptoticLowV}),
we have $\lim_{\nu\to 0}\tilde{\varepsilon}(\nu)=1/\sqrt{\Lambda}$, or equivalently 
\begin{equation}
\lim_{\nu\to 0}w(\nu)= w_{0}.
\end{equation}
Comparing the first and second terms on the right-hand side of Eq.~(\ref{eq:sup:thm:AsymptoticLowV}),
we can estimate the range of $\nu$ in which $w(\nu)\simeq w_{0}$ holds:
if $1/\sqrt{\Lambda} \gg  \nu (\sqrt{\Lambda}-1)/\Lambda$,
i.e., 
\begin{equation}
\nu\equiv\frac{V}{V_{0}}\ll \frac{\sqrt{\Lambda}}{\sqrt{\Lambda}-1}
\label{eq:sup:VforLowVelosicyRegime}
\end{equation}
is satisfied, then $w(\nu)\simeq w_{0}$
(Here, we have omitted the factor $2$ in the second term on the right-hand side of Eq.~(\ref{eq:sup:thm:AsymptoticLowV})).
When $\sqrt{\Lambda} \gg 1$, 
the condition~(\ref{eq:sup:VforLowVelosicyRegime}) is simplified as
\begin{equation}
V\ll V_{0}.
\end{equation}

\begin{figure}[t!]
\centering
\includegraphics[keepaspectratio, scale=1.0]{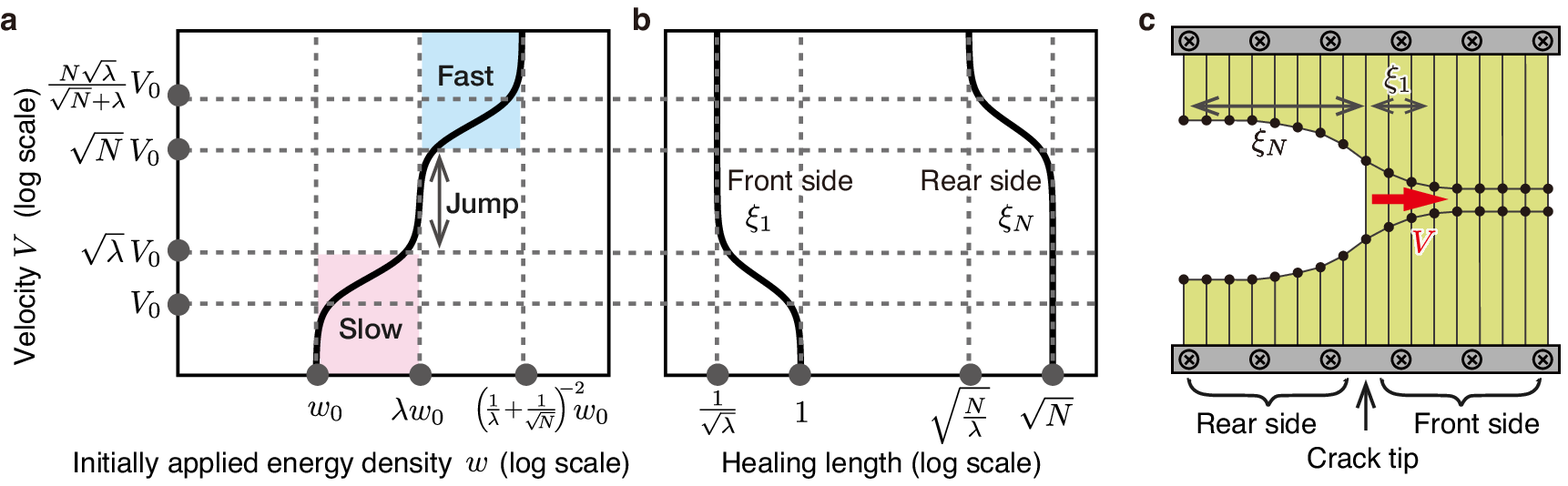}  
\caption{
\textbf{
Representative plots for typical elastomers ($\lambda=10^3$, $\Lambda=10^9$).}
\textbf{a,} $w$ vs. $V$; the velocity jump appears because the condition $\lambda \ll \Lambda$ is satisfied. 
\textbf{b,} $\xi_1$ and $\xi_\Lambda$ vs. $V$. 
The four velocity scales in plot~(a) correspond to crossover points among Zener elements in the rear and front sides of the crack tip shown in (b). 
\textbf{c,} Schematics of the healing lengths $\xi_{1}$ and $\xi_{\Lambda}$, that is, distances over which the crack shape on the rear side and the disturbance of the stress and strain distributions (mainly on the front side) recover to the remote values (see Eq.~(II.21)); as seen in (b), these healing lengths play the roles of order parameters in the context of conventional phase transitions such as superconductors and Bose-Einstein condensations \cite{S_Leggett}.
%
}
\label{fig:sup:HealingLength}
\end{figure}

Similarly, Eq.~(\ref{eq:sup:thm:AsymptoticHighV}) enables us to evaluate the $w$-$V$ relation in the high-velocity regime.
From the first term on the right-hand side of Eq.~(\ref{eq:sup:thm:AsymptoticHighV}),
we have $\lim_{\nu\to \infty}\tilde{\varepsilon} (\nu) =\lambda /(\sqrt{\Lambda} +\lambda-1)$, or equivalently 
\begin{equation}
\lim_{\nu\to \infty} w(\nu)= \frac{\lambda^2 \Lambda w_{0}}{\left(\sqrt{\Lambda} +\lambda-1\right)^2}.
\label{eq:sup:wHighVLimit}
\end{equation}
We can estimate the range of $\nu$ in which 
$w(\nu)\simeq \lambda^2 \Lambda w_{0}/(\sqrt{\Lambda} +\lambda-1)^2$ holds,
following the manner we employed in the low-velocity case.
However, we should pay attention to the fact that Eq.~(\ref{eq:sup:thm:AsymptoticHighV}) is insufficient for the estimation in the case of (unrealistic) large $\lambda$.
In fact, for example, in the Kelvin-Voigt limit ($\lambda \to \infty$) the second term on the right-hand side of Eq.~(\ref{eq:sup:thm:AsymptoticHighV}) goes to zero.
Thus, in this discussion, 
we assume $\lambda \lesssim \Lambda$,
which includes the case of typical viscoelastic materials
($1 \ll \lambda \ll \Lambda$).
Comparing the first and second terms on the right-hand side of Eq.~(\ref{eq:sup:thm:AsymptoticHighV}),
we have  the range of $\nu$ in which 
$w(\nu)\simeq \lambda^2 \Lambda w_{0}/(\sqrt{\Lambda} +\lambda-1)^2$ holds:
if the crack propagation velocity is sufficiently high, i.e., 
\begin{equation}
\nu\gg\frac{ \left(  \sqrt{\Lambda}+2\right)  \left(  \sqrt{\Lambda}-1\right)
\left(  \lambda-1\right)  ^{2} } { \left(  \sqrt{\Lambda} + \lambda-1\right) \lambda^{3/2} }
\label{eq:sup:high-vel-condition}
\end{equation}
is satisfied, then
$w(\nu)\simeq \lambda^2 \Lambda w_{0}/(\sqrt{\Lambda} +\lambda-1)^2$.
In the case of $\sqrt{\Lambda} \gg 1$ and $\lambda \gg 1$, 
which includes the case of typical viscoelastic materials, 
the above statement can be simplified, i.e., if
\begin{equation}
\nu\gg\frac{\lambda^{1/2}  \Lambda } 
{ \sqrt{\Lambda} + \lambda }
\end{equation}
is satisfied, then
$\tilde{\varepsilon} (\nu) \simeq\frac{\lambda}{\sqrt{\Lambda} +\lambda}$, or equivalently 
\begin{equation}
w(\nu)\simeq \left( \frac{1}{\sqrt{\Lambda}} + \frac{1}{\lambda} \right)^{-2} w_{0}.
\end{equation}
We summarize characteristic scales derived from Lemma~2 and Theorem~2 for $1 \ll \lambda \ll \Lambda$ in Fig.~\ref{fig:sup:HealingLength},
together with the scales associated with the jump derived in the next subsection.


\subsection{Existence condition of the velocity jump}

From Theorem~1, with the aid of Lemma~2, we can derive
the existence condition of the velocity jump as in the following theorem.\\

\textbf{Theorem~3 (Existence condition of the velocity jump).} 
\textit{
If $1<\lambda <\infty$ and
\begin{equation}
\frac{\lambda-1}{\sqrt{\lambda}}\ll\sqrt{\Lambda}%
,\label{ineq:sup:PT-condition}%
\end{equation}
then the approximate expression
\begin{equation}
\tilde{\varepsilon}(\nu)\equiv\frac{\varepsilon(\nu)}{\varepsilon_{c}}%
\simeq\sqrt{\frac{\lambda}{\Lambda}}%
\end{equation}
is valid in the range of $\nu$,
}
\begin{equation}
\frac{\lambda-1}{\sqrt{\lambda}}\ll\nu\ll\sqrt{\Lambda}%
.\label{ineq:sup:PT-region}%
\end{equation}\\



\noindent
\textit{Proof of Theorem~3.}

We evaluate the positive solution of the characteristic equation~(\ref{eq:sup:def:gLambda}) 
under the condition~(\ref{ineq:sup:PT-region}). 
According to Lemma~2, the positive solution of the characteristic equation
$g_{\Lambda}(\xi)=0$ is expressed as $\xi\simeq\xi_{\Lambda}^{(0)}\equiv\sqrt{\Lambda}$
for $\nu\ll\sqrt{\Lambda}$.
Similarly, 
the positive solution of $g_{1}(\xi)=0$ is expressed as $\xi\simeq\xi_{1}^{(\infty)}\equiv1/\sqrt{\lambda}$
for $(\lambda-1)^{2}/\lambda^{3/2}<(\lambda-1)/\lambda^{1/2}\ll\nu$. 
Then, Eq.~(\ref{eq:sup:thm:epsilonnu}) can be expressed as
\begin{equation}
\tilde{\varepsilon}
\equiv\frac{\varepsilon}{\varepsilon_{c}}
\simeq
\frac{\frac{\nu}{\lambda-1}\left(\frac{\Lambda}{\xi_{1}^{(\infty)}\xi_{\Lambda}^{(0)}}+\lambda\right)
                +\xi_{1}^{(\infty)}+\xi_{\Lambda}^{(0)}}
         {\frac{\nu\Lambda}{\lambda-1}\left( 1+\frac{1}{\xi_{1}^{(\infty)}\xi_{\Lambda}^{(0)}}\right)
                +\nu +\Lambda\xi_{1}^{(\infty)}+\xi_{\Lambda}^{(0)}}
=\frac{\frac{\nu}{\lambda-1}\left(  \lambda+\sqrt{\lambda\Lambda}\right)  +  \frac{1}{\sqrt{\lambda}}+\sqrt{\Lambda}}
         {\frac{\nu}{\lambda-1}\left(\Lambda+\sqrt{\lambda\Lambda}\right)  +\nu
                 +\frac{\Lambda}{\sqrt{\lambda}}+\sqrt{\Lambda}}.
\label{eq:sup:e_at_transitionpoint_proof}
\end{equation}
By noting the relation,
$\frac{1}{\sqrt{\lambda}}+\sqrt{\Lambda}
<   \sqrt{\lambda}+\sqrt{\Lambda}
\ll   \frac{\nu}{\lambda-1}\left(  \lambda+\sqrt{\lambda\Lambda}\right)$,
for the numerator in Eq.~(\ref{eq:sup:e_at_transitionpoint_proof}),
and the relation,
\begin{equation}
\frac{\Lambda}{\sqrt{\lambda}}+\sqrt{\Lambda}
\ll
\frac{\nu\Lambda}{\lambda-1}\left(  1+\sqrt{\frac{\lambda}{\Lambda}}\right),
\end{equation}
for the denominator in Eq.~(\ref{eq:sup:e_at_transitionpoint_proof}),
we obtain
\begin{equation}
\tilde{\varepsilon}\simeq\frac{\sqrt{\lambda\Lambda}+\lambda}{\Lambda
+\sqrt{\lambda\Lambda}+\lambda-1}\simeq\frac{\sqrt{\lambda\Lambda}+\lambda
}{\Lambda+\sqrt{\lambda\Lambda}}=\sqrt{\frac{\lambda}{\Lambda}}.
\end{equation}
(Q.E.D.)\\

We have two remarks for Theorem~3:
(i) the present model consisting of Kelvin-Voigt elements (which corresponds
to $\lambda\rightarrow\infty$) never satisfies the existence condition of
velocity jump in Eq.~(\ref{ineq:sup:PT-condition}) because $\Lambda$ is finite;
(ii) if $\lambda\gg1$, then the existence
condition~(\ref{ineq:sup:PT-condition}) reduces to 
\begin{equation}
\sqrt{\lambda}\ll\sqrt{\Lambda}.
\end{equation}

\bibliographystyle{apsrev}

\end{document}